\documentclass[useAMS,usenatbib]{mnras}

\usepackage{revsymb}
\usepackage{amsmath}
\usepackage{amsfonts}
\usepackage{amssymb}
\usepackage{lastpage}
\usepackage{graphicx}
\usepackage{multicol}

\title[Aberration in the cosmic microwave background radiation]
      {Determining our peculiar velocity from the aberration in the
       cosmic microwave background}
\author[R. Aurich and D. Reinhardt]
  {R.~Aurich and D.~Reinhardt \\
  Institut f\"ur Theoretische Physik, Universit\"at Ulm,\\
  Albert-Einstein-Allee 11,\\ D-89069 Ulm, Germany
}

\begin{document}

\date{}

\pagerange{\pageref{firstpage}--\pageref{LastPage}} \pubyear{2021}

\def\LaTeX{L\kern-.36em\raise.3ex\hbox{a}\kern-.15em
    T\kern-.1667em\lower.7ex\hbox{E}\kern-.125emX}

\newtheorem{theorem}{Theorem}[section]

\def\bfis{\hbox{\scriptsize\rm i}}
\def\bfi{\hbox{\rm i}}
\def\bfj{\hbox{\rm j}}

\setlength{\topmargin}{-1cm}

\label{firstpage}

\maketitle

\begin{abstract}
  The motion of our solar system relative to the cosmic microwave background
  (CMB) rest frame leads to
  subtle distortions in the observed CMB sky map due to the aberration effect.
  Usually the corresponding peculiar velocity is determined from the
  CMB dipole but neglecting intrinsic dipole contributions.
  Here it is investigated whether certain invariant scalar measures,
  which are derived from first and second order covariant derivatives
  on the sphere,
  can detect the distortions caused by the aberration effect at high multipoles.
  This would in principle allow to disentangle the Doppler from intrinsic dipole
  contributions providing an independent method for the determination
  of our peculiar velocity.
  It is found that the eigenvalues of the Hessian matrix of the
  temperature field are well suited for that task.
\end{abstract}

\begin{keywords}
Cosmology, cosmic background radiation
\end{keywords}


\section{Introduction}

On the theoretical side the fluctuations in the cosmic microwave background
(CMB) radiation are usually computed in the rest frame of the CMB with
computer programs such as CAMB.
However, on the observational side, one has to take into account our
peculiar velocity with respect to the rest frame of the CMB.
This leads to the large Doppler dipole contribution in the CMB
on which the usual determination of our velocity is based.
For further cosmological CMB analysis this contribution is ignored.

The peculiar velocity is defined as the velocity of our
solar system barycentre relative to the CMB rest frame,
which in turn is defined by the vanishing of the temperature dipole $a_{1m}$.
These coefficients are extracted from the spherical expansion of the
temperature fluctuations
\begin{equation}
  \label{eq:spherical_expansion}
  \delta T(\vartheta,\varphi) \; = \;
  \sum_{\ell,m} a_{\ell m} \, Y_{\ell m}(\vartheta,\varphi)
  \hspace{10pt} ,
\end{equation}
where $Y_{\ell m}(\vartheta,\varphi)$ are the spherical harmonics.
The COBE mission determined the peculiar velocity as $\beta_\odot = 0.00123$
in the direction $(l,b)\simeq(264^\circ, 48^\circ)$ in galactic coordinates
\citep{Kogut_et_al_1993},
where $\beta:= v/c$ is the velocity in units of the speed of light $c$.

If $T^\text{(rest)}$ and $\hat n^\text{(rest)}$ denote the CMB rest frame
temperature and the unit vector of the considered direction, respectively,
then a Lorentz transformation according to the peculiar velocity $\vec\beta$
leads to the transformed temperature \citep{Challinor_Leeuwen_2002}
\begin{equation}
  \label{eq:temperature_transformation}
  T^\text{(obs)}(\hat n^\text{(obs)}) \; = \;
  \frac{T^\text{(rest)}(\hat n^\text{(rest)})}
  {\gamma(1-\hat n^\text{(obs)}\cdot\vec\beta\,)}
\end{equation}
in the frame of the observer,
who measures $T^\text{(obs)}(\hat n^\text{(obs)})$ in the
direction $\hat n^\text{(obs)}$ and $\gamma := (1-\beta^2)^{-1/2}$.
It is instructive to consider this expression in linear order of $\beta$,
which leads to the temperature fluctuations
(see e.\,g.\ \cite{Planck_Doppler_Boosting_2013})
\begin{eqnarray}\nonumber
  \delta T^\text{(obs)}(\hat n^\text{(obs)}) & = &
  T_0 \, \hat n^\text{(obs)}\cdot\vec\beta 
\\ &  & \hspace*{-70pt} + \;
  \label{eq:temperature_fluctuations_beta}
  \delta T^\text{(rest)}(\hat n^\text{(obs)}-
                        \nabla\hat n^\text{(obs)}\cdot\vec\beta\,) \;
         (1+\hat n^\text{(obs)}\cdot\vec\beta\,)
  \hspace{5pt} ,
\end{eqnarray}
where $T_0$ denotes the CMB mean temperature.
The first term on the right hand side of
eq.\,(\ref{eq:temperature_fluctuations_beta}) is the usual Doppler dipole
which has an amplitude of $T_0\beta_\odot\simeq 3352 \mu\hbox{K}$.
The argument of the temperature fluctuation $\delta T^\text{(rest)}$ in
eq.\,(\ref{eq:temperature_fluctuations_beta})
is aberrated by the deflection $\nabla\hat n^\text{(obs)}\cdot\vec\beta$.
Furthermore, this amplitude is also modulated by the factor
$(1+\hat n^\text{(obs)}\cdot\vec\beta\,)$.
The numerical analysis in this work is based on the complete formula
(\ref{eq:temperature_transformation}) and not on the linear approximation.
The angular power spectrum ${\cal D}_\ell = \ell(\ell+1) C_\ell/(2\pi)$
with $C_\ell = \sum_m |a_{\ell m}|^2/(2\ell+1)$ possesses for small
multipole orders $2<\ell \lesssim 10$ values around $1000 \mu\hbox{K}^2$
\citep{Planck_2018_VII}.
Assuming that the intrinsic dipole ${\cal D}_1$ is of the same order,
one is lead to the expectation
$C_1 = \pi {\cal D}_1 \simeq \pi\, 1000 \mu\hbox{K}^2$
and thus to a dipole amplitude $\sqrt{C_1} \simeq 56\mu\hbox{K}$.
An intrinsic dipole of the primordial temperature anisotropy field is therefore
expected to be at the 1\% level with respect to the Doppler dipole
$T_0\beta_\odot$.

Although intrinsic dipole of the primordial temperature anisotropy field is
considered as negligible,
the observed CMB power spectrum reveals a parity asymmetry
and possesses multipoles $C_\ell$
with enhanced power for odd multipoles $\ell$ and
reduced power for the even ones for not too high values of $\ell$ corresponding
to the largest angular scales
\citep{Land_Magueijo_2005,Kim_Naselsky_2010,Planck_2018_VII}.
And since the dipole is an odd multipole, of course, it could be larger
than the expected 1\% level.

Due to the inability to disentangle the Doppler contribution from other
dipole contributions,
it would thus be favourable to derive the peculiar velocity from
an alternative way.
This is offered by the fact that the aberration due to the peculiar velocity
induces a stretching and a compression of the CMB structures
in sky maps, which depends on the angular distance from the direction of our
peculiar velocity,
in addition to the modulation described above.
These modifications of the CMB structures occur for a velocity $\beta$
of the order $10^{-3}$ at very high multipoles $\ell\gtrsim 1000$.
Based on eq.\,(\ref{eq:temperature_transformation}),
they are described by the aberration kernel $K_{\ell m}^{\ell'm'}(\beta)$
\citep{Challinor_Leeuwen_2002}
of the multipole transformation law
\begin{equation}
  \label{eq:aberration_kernel_double}
  a^\text{(obs)}_{\ell m} \; = \;
  \sum_{\ell',m'} K_{\ell m}^{\ell'm'}(\beta) \; a^\text{(rest)}_{\ell'm'}
  \hspace{10pt} ,
\end{equation}  
where $a^\text{(rest)}_{\ell m}$ are the multipole expansion coefficients
in the CMB rest frame
and $a^\text{(obs)}_{\ell m}$ are those observed in the moving frame.
The double sum in eq.\,(\ref{eq:aberration_kernel_double}) reduces to a single
sum by aligning the $z$-axis of the spherical coordinate system with the
direction of the velocity
\begin{equation}
  \label{eq:aberration_kernel_z}
  a^\text{(obs)}_{\ell m} \; = \;
  \sum_{\ell'} K_{\ell m}^{\ell'm}(\beta) \; a^\text{(rest)}_{\ell'm}
  \hspace{10pt} ,
\end{equation}  
which simplifies the computations drastically.
This is due to the fact that in this special case the aberration causes
angular distortions only in the direction of the polar angle $\vartheta$.
The aberration kernel in eq.\,(\ref{eq:aberration_kernel_z}) can be computed by
the recursion algorithm of \cite{Chluba_2011},
see also \citep{Dai_Chluba_2014}.

The multipole transformation leads to a coupling between nearby multipoles.
Aligning the $z$-axis of the spherical coordinate system with
the boost in the forward direction,
the temperature fluctuation structures are compressed towards the north pole
leading to a power leakage from $C_\ell$ to higher multipoles
$C_{\ell+1},C_{\ell+2},\dots$ on the northern hemisphere.
On the southern hemisphere, the power leakage occurs towards lower multipoles
$C_{\ell-1},C_{\ell-2},\dots$ since the structures are stretched there.
In full sky analyses these power redistributions partly cancel, and
in partial sky analyses, special care has to be taken into account
\citep{Catena_Notari_2013}.
Methods to detect these mode couplings between neighbouring multipoles
were suggested by \cite{Burles_Rappaport_2006,Kosowsky_Kahniashvili_2011} and
\cite{Amendola_et_al_2011}.
The \cite{Planck_Doppler_Boosting_2013} finds the mode couplings to be
consistent with the peculiar velocity determined by the Doppler dipole.
The dependence of the magnitude of the power leakage on the angular distance
from the boost direction is analysed by \cite{Jeong_et_al_2014} and discussed
with respect to hemispherical asymmetries by \cite{Notari_Quartin_Catena_2014}.
These considerations are extended to the polarization measurements of the CMB
in \citep{Amendola_et_al_2011,Dai_Chluba_2014,Mukherjee_De_Souradeep_2014,Yasini_Pierpaoli_2017,Yasini_Pierpaoli_2020}.

In real sky observations, one has always to deal with a masked sky
which also leads to a power leakage in neighbouring multipoles
(see e.\,g.\ \cite{Catena_Notari_2013}).
Therefore, the so-called pseudo-$C_\ell$'s possess a mode coupling due to
the aberration effect and to the partial sky observations.
So one might alternatively turn to quantities
that can be derived from local CMB data without the necessity to extract
pseudo-$a_{\ell m}$'s obtained from the harmonic analysis on the full sky.
This can be done by considering quantities derived from the first and second
covariant derivatives on the sphere with respect to $\vartheta$
and $\varphi$.
These derivatives, however, depend on the chosen $z$-axis of the
spherical coordinate system.
Using these derivatives directly would be a drawback analogously to some methods
employing strength of the mode coupling with respect to the angular distance
from the boost direction.
It is much more favourable to construct from the covariant derivatives
with respect to $\vartheta$ and $\varphi$ scalar quantities
which are independent of the chosen $z$-axis.
Out of the many possibilities,
\cite{Monteserin_Barreiro_Sanz_Martinez-Gonzalez_2005} suggest
twelve such scalar quantities suitable for general applications
to the cosmic microwave background.
All of them are based on the first and second covariant derivatives of
the temperature field $\delta T(\vartheta,\varphi)$.
\cite{Dore_Colombi_Bouchet_2003} discuss the geometrical meaning of the
local curvature structure in the form as 'hill', 'lake' and 'saddle' regions.

It is the aim of this work to investigate which scalar statistics
on the sphere suggested by
\cite{Monteserin_Barreiro_Sanz_Martinez-Gonzalez_2005}
are useful as an alternative method for the determination of our
peculiar velocity.

\begin{figure*}
\begin{multicols}{2}
  \includegraphics[width=\linewidth]{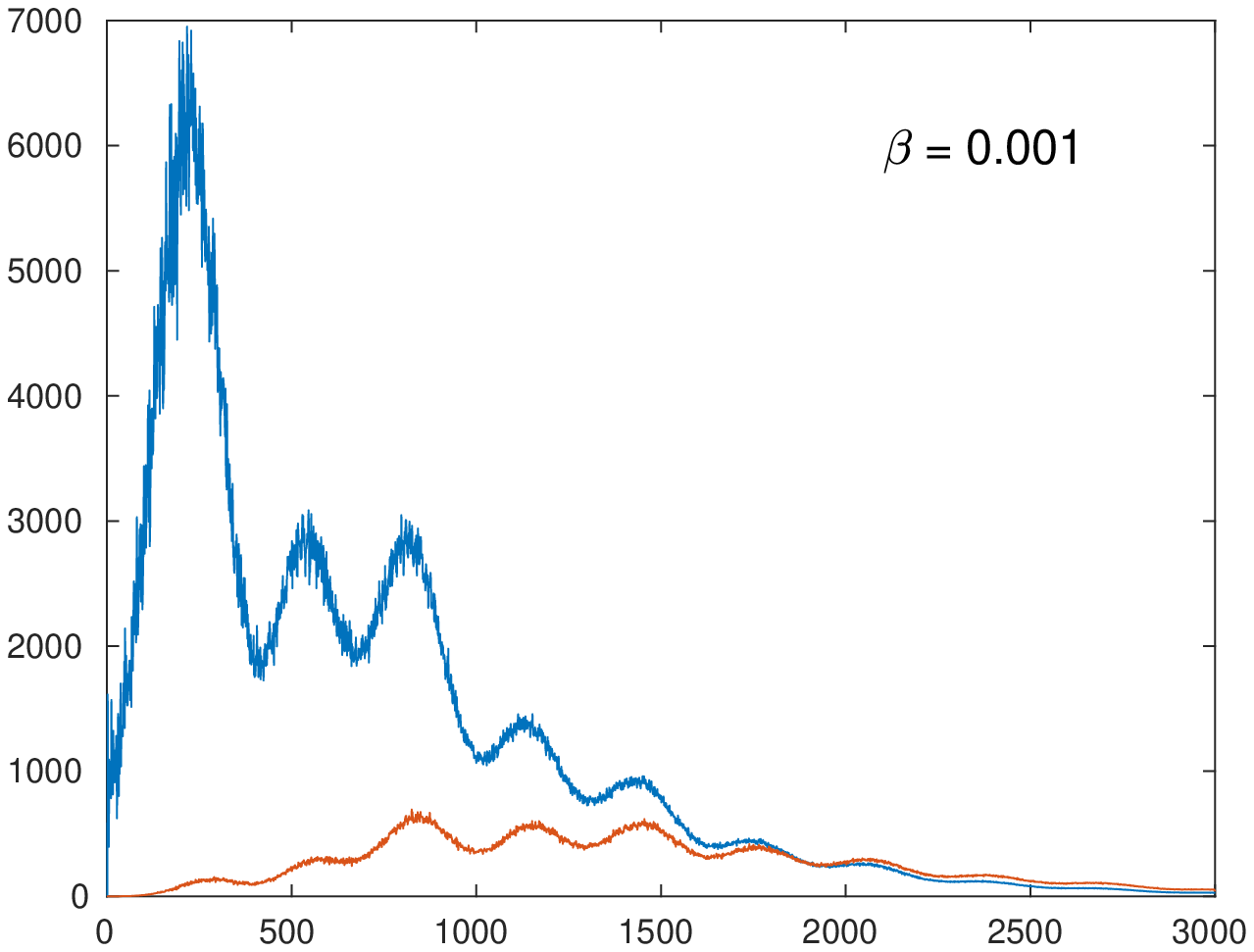}
  \put(-180,150){(a)}
  \put(-40,11){$\ell$}
  \put(-230,155){${\cal D}_\ell$}
  \put(-235,133){$[\mu$K$^2]$}
  \par 
  \includegraphics[width=\linewidth]{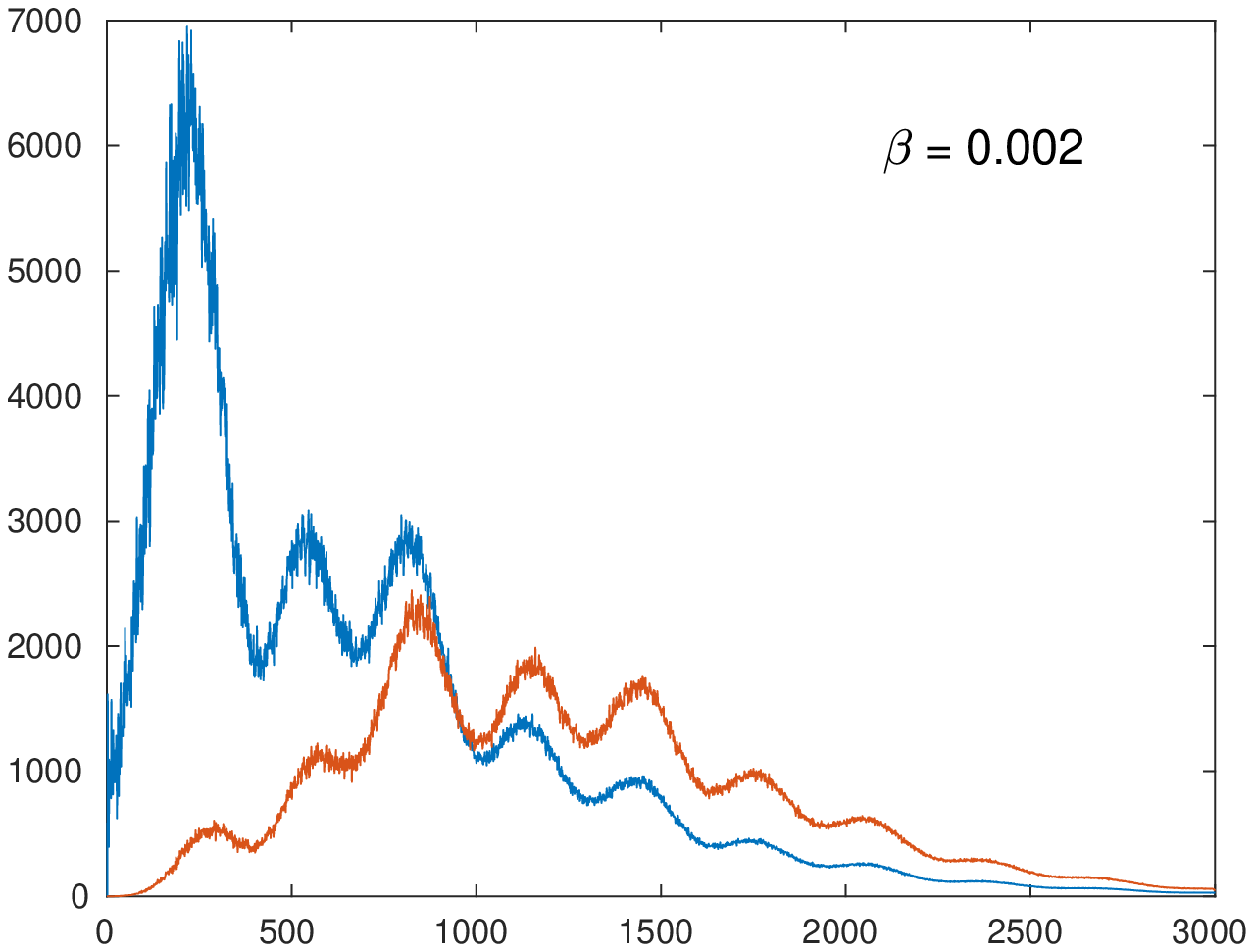}
  \put(-180,150){(b)}
  \put(-40,11){$\ell$}
  \put(-230,155){${\cal D}_\ell$}
  \put(-235,133){$[\mu$K$^2]$}
  \par
\end{multicols}
\begin{multicols}{2}
  \includegraphics[width=\linewidth]{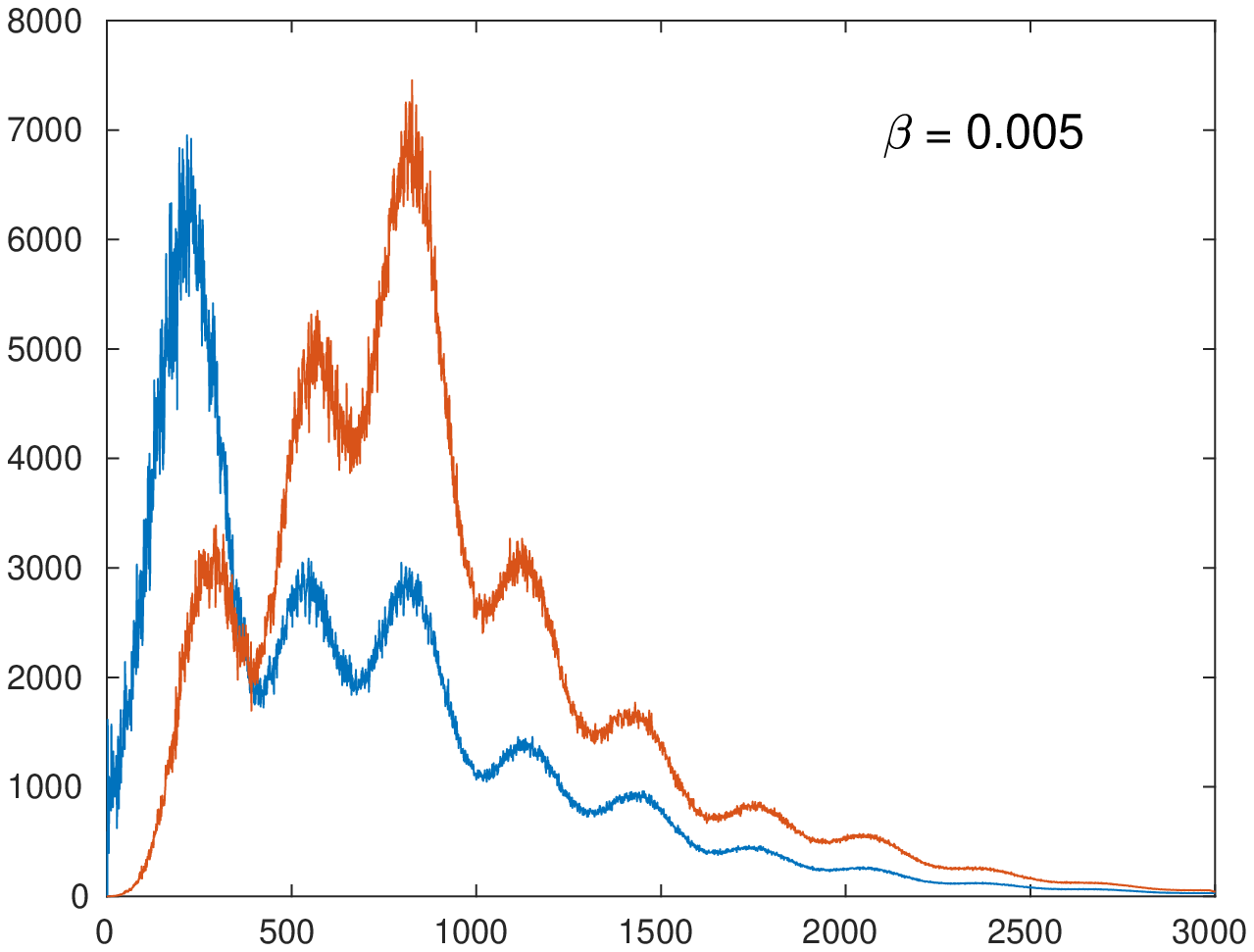}
  \put(-180,150){(c)}
  \put(-40,11){$\ell$}
  \put(-230,155){${\cal D}_\ell$}
  \put(-235,137){$[\mu$K$^2]$}
  \par 
  \includegraphics[width=\linewidth]{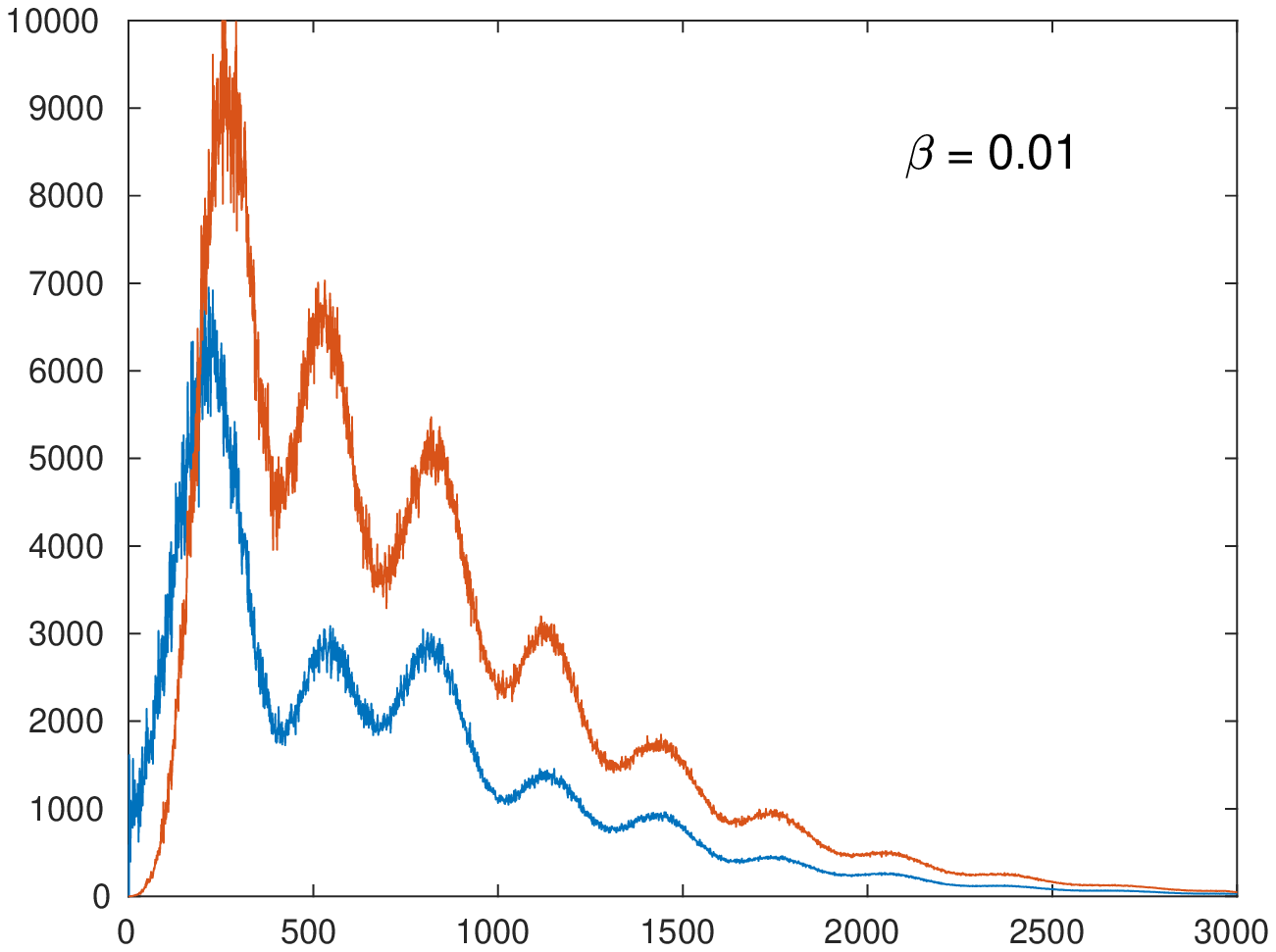}
  \put(-180,150){(d)}
  \put(-40,11){$\ell$}
  \put(-235,144){${\cal D}_\ell$}
  \put(-240,128){$[\mu$K$^2]$}
  \par
\end{multicols}
\caption{\label{Fig:angular_power_spectra}
  The multipole shifting spectra
  $\Delta {\cal D}_\ell = \ell(\ell+1) \Delta C_\ell/(2\pi)$ with $\Delta C_\ell$
  defined in eq.\,(\ref{eq:multipole_shifting_spectra}) are plotted in red
  for four different peculiar velocities $\beta$.
  In addition, the corresponding CMB rest frame angular power spectrum
  ${\cal D}_\ell = \ell(\ell+1) C^\text{(rest)}_\ell/(2\pi)$ is shown as
  the same blue curve in all four panels in order to allow a comparison.
  All four cases are based on the same CMB rest frame realization.
}
\end{figure*}


\section{Scalar Measures on the Sphere}
\label{Sec:Scalars_Sphere}

The first covariant derivatives of the temperature field
$\delta T(\vartheta,\varphi)$ are the usual partial derivatives
denoted by a comma.
As usual, the covariant derivatives are denoted by semicolons.
The second covariant derivatives
\begin{equation}
  \label{eq:covariant_derivative}
  \delta T_{;ij} \; = \;
  \delta T_{,ij} \, - \, \Gamma_{ij}^k \, \delta T_{,k}
  \hspace{10pt} , \hspace{10pt}
  i,j,k \in \{\vartheta,\varphi\}
  \hspace{10pt} ,
\end{equation}  
require the Christoffel symbols $\Gamma_{ij}^k$
which are non-vanishing only for
$(ds^2 = d\vartheta^2 + \sin^2\vartheta\, d\varphi^2)$
$$
  \Gamma_{\varphi\vartheta}^\varphi = \Gamma_{\vartheta\varphi}^\varphi =
  \frac{\cos\vartheta}{\sin\vartheta}
  \hspace{10pt} \hbox{ and } \hspace{10pt}
  \Gamma_{\varphi\varphi}^\vartheta = -\sin\vartheta \, \cos\vartheta
  \hspace{10pt} .
$$
The twelve scalars suggested by
\cite{Monteserin_Barreiro_Sanz_Martinez-Gonzalez_2005}
can be classified into the following groups:
\begin{itemize}
\item[(i)]
  Scalars derived from the Hessian matrix\\
  The negative Hessian matrix $-{\cal H}=(-\delta T_{;ij})$ has as interesting
  quantities the eigenvalues, the trace and the determinant.
  The eigenvalues $\lambda_1$ and $\lambda_2$ can be expressed by the
  covariant derivatives as
 \begin{equation}
  \label{eq:lambda_1}
  \lambda_1 \, = \, -\frac 12 \delta T^{\;\,;i}_{;i}
  + \frac 12 \sqrt{(\delta T^{\;\,;i}_{;i})^2 -
    2(\delta T^{\;\,;i}_{;i}\delta T^{\;\,;j}_{;j}-
      \delta T^{\;\,;j}_{;i}\delta T^{\;\,;i}_{;j})}
 \end{equation}
 and
 \begin{equation}
  \label{eq:lambda_2}
  \lambda_2 \, = \, -\frac 12 \delta T^{\;\,;i}_{;i}
  - \frac 12 \sqrt{(\delta T^{\;\,;i}_{;i})^2 -
    2(\delta T^{\;\,;i}_{;i}\delta T^{\;\,;j}_{;j}-
      \delta T^{\;\,;j}_{;i}\delta T^{\;\,;i}_{;j})} \; .
 \end{equation}
 The trace of the Hessian matrix ${\cal H}$ corresponds also to the Laplacian
 and can be expressed by the eigenvalues
 \begin{equation}
  \label{eq:H_trace}
  \lambda_+ \, = \, \hbox{tr }{\cal H}  \, = \, \delta T^{\;\,;i}_{;i} \, = \,
  -\lambda_1 \, - \, \lambda_2
  \hspace{10pt} .
 \end{equation}
 And finally, the determinant of the negative Hessian matrix is
 \begin{equation}
  \label{eq:H_det}
  d \, = \, -\hbox{det }{\cal H}  \, = \,
  \frac 12\left[\delta T^{\;\,;i}_{;i}\delta T^{\;\,;j}_{;j} -
                \delta T^{\;\,;j}_{;i}\delta T^{\;\,;i}_{;j} \right] \, = \,
  \lambda_1 \, \lambda_2
  \hspace{10pt} .
 \end{equation}
 
\item[(ii)]
 Distortion scalars\\
 The four scalars, which are chosen to characterize the distortion of the
 temperature field $\delta T(\vartheta,\varphi)$, are the shear,
 the distortion, the ellipticity and the shape index.
 They are defined as follow.
 The shear $y$ is given as
 \begin{eqnarray}
  y & = & \nonumber
  \frac 14\left[\delta T^{\;\,;i}_{;i} \right]^2 -
  \frac 12\left[\delta T^{\;\,;i}_{;i}\delta T^{\;\,;j}_{;j} -
                \delta T^{\;\,;j}_{;i}\delta T^{\;\,;i}_{;j} \right] \\
  & = &
  \frac 14\,(\lambda_1 - \lambda_2)^2 \; ,
  \label{eq:shear}
 \end{eqnarray}
 the distortion $\lambda_-$ as
 \begin{equation}
  \label{eq:distortion}
  \lambda_- \, = \,
  \sqrt{\left[\delta T^{\;\,;i}_{;i} \right]^2 -
         2\left[\delta T^{\;\,;i}_{;i}\delta T^{\;\,;j}_{;j} -
                \delta T^{\;\,;j}_{;i}\delta T^{\;\,;i}_{;j} \right]} \, = \,
  \lambda_1 - \lambda_2 \; ,
 \end{equation}
 the ellipticity $e$ as
 \begin{eqnarray}
  e & = & \nonumber
  -\frac 1{2\left[\delta T^{\;\,;i}_{;i} \right]} \,
  \sqrt{\left[\delta T^{\;\,;i}_{;i} \right]^2 -
         2\left[\delta T^{\;\,;i}_{;i}\delta T^{\;\,;j}_{;j} -
                \delta T^{\;\,;j}_{;i}\delta T^{\;\,;i}_{;j} \right]} \\
  & = &
  \frac 12\,\frac{\lambda_1 - \lambda_2}{\lambda_1 + \lambda_2} \; ,
  \label{eq:ellipticity}
 \end{eqnarray}
 and the shape index $\iota$ as
  \begin{equation}
  \label{eq:shape_index}
  \iota \, = \, \frac 2\pi \, \hbox{ arctan}\left(-\frac{1}{2e}\right)
  \hspace{10pt} .
 \end{equation}
\item[(iii)]
 Scalars derived from the gradient\\
 The squared modulus of the gradient field and its derivative are considered
 here to extract the level of smoothness of the temperature field.
 The square of the gradient modulus $g$ is given as
 \begin{equation}
  \label{eq:gradient_g}
  g \, = \, \delta T^{,i}\, \delta T_{,i}
  \hspace{10pt} ,
 \end{equation}
 and a derivative of it as
 \begin{equation}
  \label{eq:gradient_Dg}
  D_g \, = \, \frac d{ds}\left(\frac 12 g \right)
  \, = \, \delta T^{;ij}\, \delta T_{,i}\, \delta T_{,j}
  \hspace{10pt} ,
 \end{equation}
 where the derivative has to be taken along the arc $s$ associated to the
 integral curve of the gradient.
 \item[(iv)]
 Curvature scalars\\
 The last two scalars considered here are the Gaussian curvature
 \begin{equation}
 \label{eq:Gaussian_curvature}
 \kappa_\text{G} \, = \, \frac 12 \,
 \frac{\delta T^{\;\,;i}_{;i} \delta T^{\;\,;j}_{;j} -
       \delta T^{\;\,;j}_{;i}\delta T^{\;\,;i}_{;j}}
       {(1 + \left[\delta T^{,i}\delta T_{,i}\right])^2}
 \end{equation}
 and the extrinsic curvature being the average of the two principal curvatures
 of the temperature field can be expressed by the quantities defined so far
 \begin{equation}
 \label{eq:extrinsic_curvature}
 \kappa_\text{ex} \, = \, \frac 12 \, \frac 1{\sqrt{1+g}}
      \left[ \lambda_+ - \frac{D_g}{1+g} \right]
 \hspace{10pt} .
 \end{equation}
\end{itemize}
For further details on these twelve scalar measures, see the discussions in
\cite{Monteserin_Barreiro_Sanz_Martinez-Gonzalez_2005}.

\section{The ability of scalar measures to detect the aberration}

All the scalar measures presented in the previous section are based on
covariant derivatives on the temperature field $\delta T(\vartheta,\varphi)$
and are independent of the chosen orientation of the spherical coordinate
system.
But they differ in the ability to detect the distortions caused by the
aberration.
In order to find the most suitable scalar measures,
these measures are computed from a set of 1000 simulated CMB sky maps
such that the deviation of the detected boost direction from that used in
the simulations allows such a decision.

The CMB simulations are based on the power spectrum
which is computed using
CAMB\footnote{The software was written by A.~Lewis and A.~Challinor
and may be downloaded from \url{http://camb.info/}}
for the cosmological parameters of the
$\Lambda$CDM concordance model given by the \cite{Planck_2018_I}
in their table 6 in the column 'Planck+BAO'.
The main parameters are $\Omega_\text{b}h^2 = 0.022447$,
$\Omega_\text{c}h^2 = 0.11923$, and $h=0.677$.
A large set of CMB sky map realizations is generated
from this power spectrum which yields the spherical expansion coefficients
$a^\text{(rest)}_{\ell m}$ in the CMB rest frame.
Then these sky maps are boosted in the $z$-direction using
eq.\,(\ref{eq:aberration_kernel_z}) for several peculiar velocities $\beta$
in order to obtain the coefficients $a^\text{(obs)}_{\ell m}$.
The aberration kernel is computed by the recursion algorithm
of \cite{Chluba_2011}.
To allow a decision on the quality of the scalar measures of section
\ref{Sec:Scalars_Sphere} free of numerical artefacts,
all computations are carried out in harmonic space up to the multipole
order of $\ell_\text{max} = 3000$,
so that the required differentiations with respect to $\vartheta$ and
$\varphi$ are also computed from the coefficients $a^\text{(obs)}_{\ell m}$.
Only in the final step, the sky maps of the scalar measures are generated
in the HEALPix format
\citep{Gorski_Hivon_Banday_Wandelt_Hansen_Reinecke_Bartelmann_2005}
with a resolution of $N_\text{side}=2048$.

The effect of the aberration can be studied in the multipole spectra $C_\ell$
by considering the relative difference
$(C_\ell^\text{(obs)}-C_\ell^\text{(rest)})/C_\ell^\text{(rest)}$
as done by \cite{Jeong_et_al_2014}
which is considering boosted values $C_\ell^\text{(obs)}$ minus the rest frame
values $C_\ell^\text{(rest)}$ in the power spectrum.
Another instructive possibility is to consider the difference in the
expansion coefficients $a_{\ell m}$ by defining the
multipole shifting spectra
\begin{equation}
\label{eq:multipole_shifting_spectra}
\Delta C_\ell \; = \;
\frac 1{2\ell+1} \,
\sum_{m=-\ell}^\ell |a^\text{(obs)}_{\ell m} - a^\text{(rest)}_{\ell m}|^2
\hspace{10pt} .
\end{equation}
The effect of boosting the CMB radiation by four different velocities $\beta$
is shown in figure \ref{Fig:angular_power_spectra}
where the multipole shifting power spectrum
$\Delta {\cal D}_\ell = \ell(\ell+1) \Delta C_\ell/(2\pi)$ with $\Delta C_\ell$
defined above in eq.\,(\ref{eq:multipole_shifting_spectra})
is plotted in comparison with the CMB rest frame angular power spectrum
${\cal D}_\ell = \ell(\ell+1) C^\text{(rest)}_\ell/(2\pi)$.
The peculiar velocity $\beta= 0.001$,
which is shown in figure \ref{Fig:angular_power_spectra}(a),
is close to our solar system value $\beta_\odot= 0.00123$ and reveals an
appreciable influence of the aberration above $\ell \gtrsim 800$.
At multipoles above $\ell \simeq 2000$ the multipole shifting power spectrum
$\Delta {\cal D}_\ell$ becomes even larger than ${\cal D}_\ell$ in this case.
For increasingly larger values of $\beta$,
the multipole shifting power spectrum $\Delta {\cal D}_\ell$ begins to
dominate at ever smaller multipole orders as seen in figures
\ref{Fig:angular_power_spectra}(b) to \ref{Fig:angular_power_spectra}(d).
However, for the most interesting case of the solar system barycentre velocity,
one has to carry out the analysis with sufficiently high multipole orders
and thus, all calculations of this work are done with a cut-off of
$\ell_\text{max}=3000$.

\begin{figure}
\begin{center}
\begin{minipage}{9cm}
\vspace*{-3pt}
\begin{minipage}{9cm}
  \hspace*{-10pt}\includegraphics[width=9.0cm]{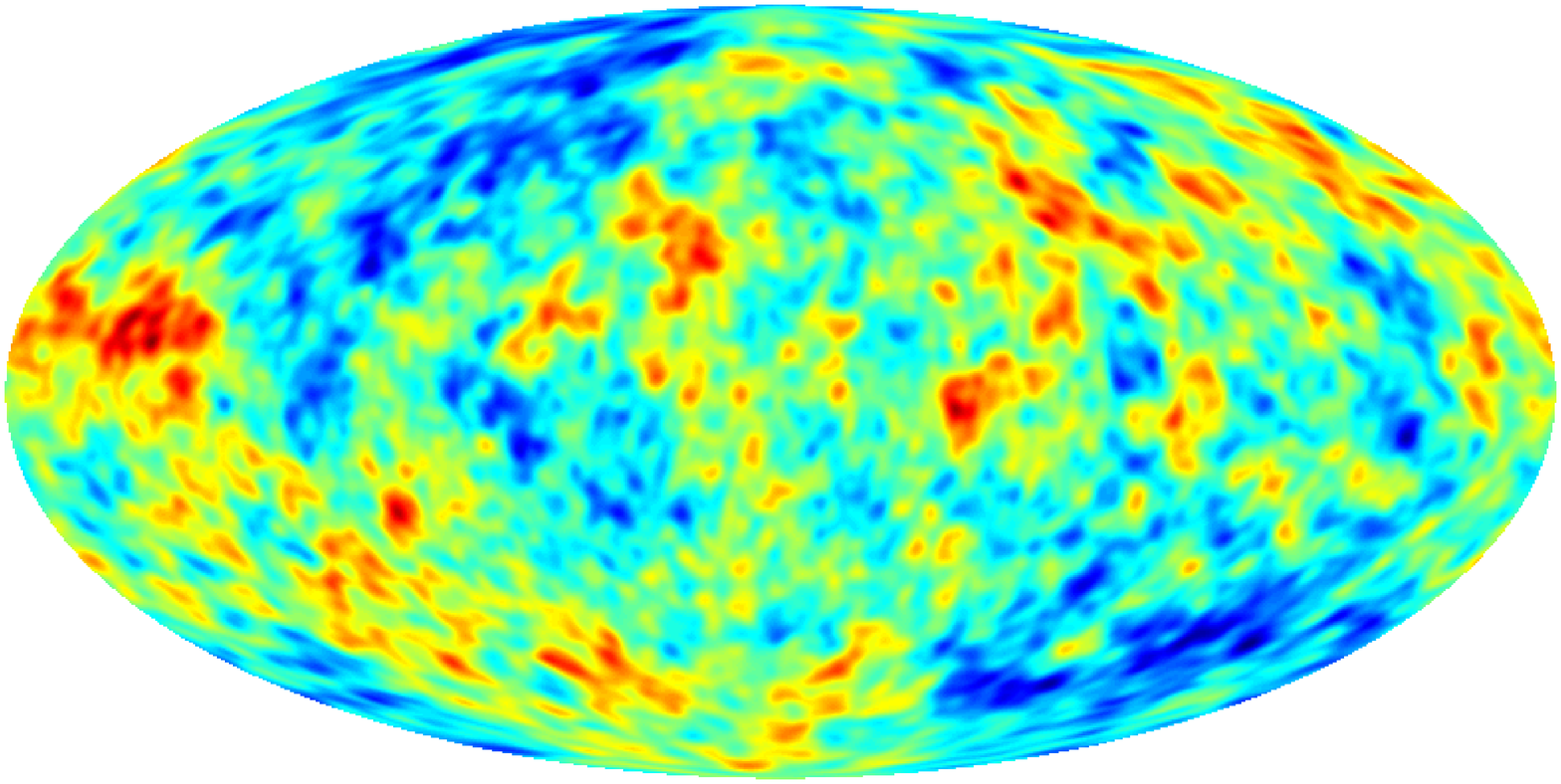}
  \put(-260,123){(a)  temperature map}
\end{minipage}
\begin{minipage}{9cm}
  \hspace*{-10pt}\includegraphics[width=9.0cm]{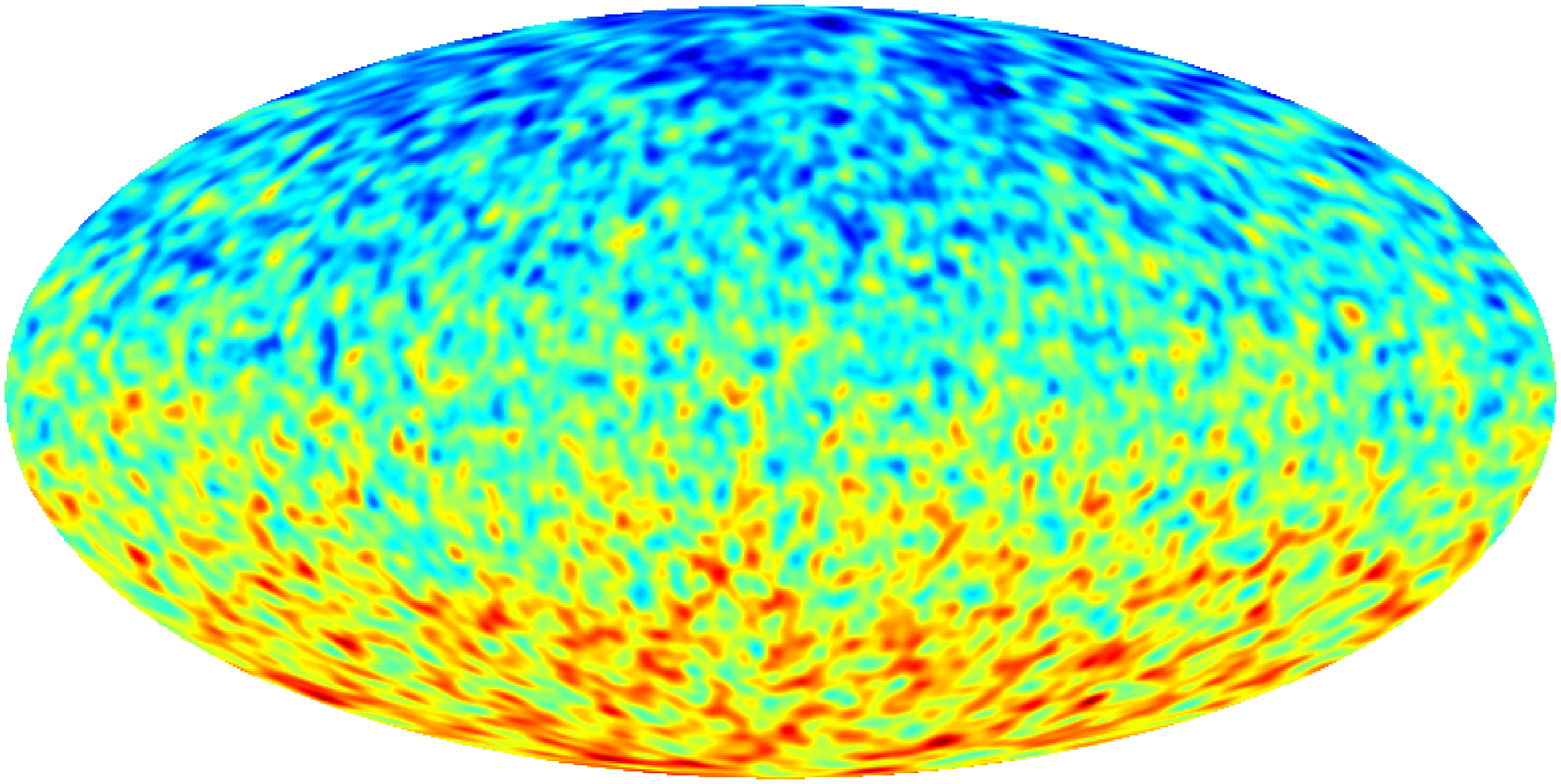}
  \put(-260,123){(b)  eigenvalue $\lambda_1$}
\end{minipage}
\begin{minipage}{9cm}
  \hspace*{-10pt}\includegraphics[width=9.0cm]{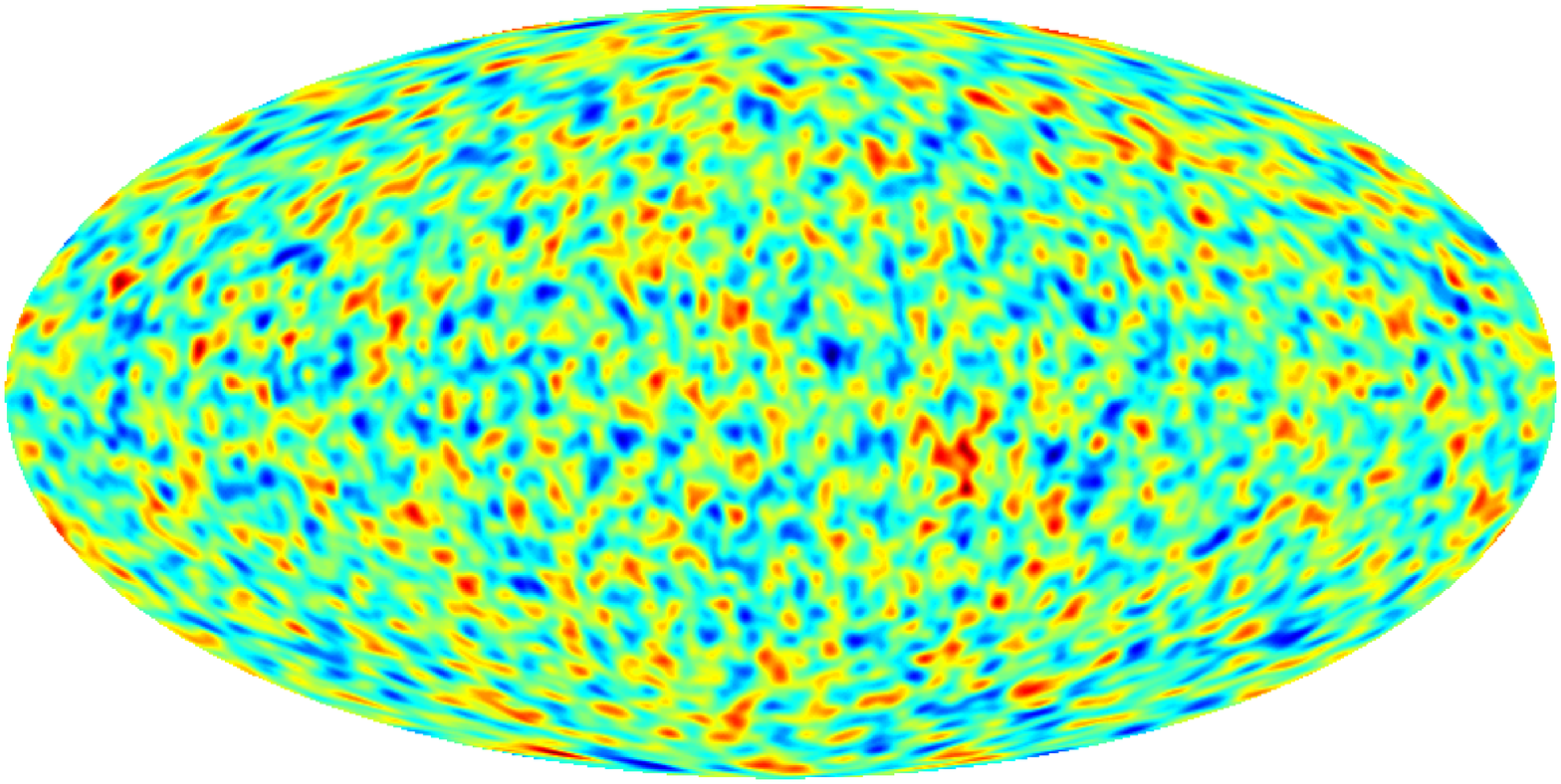}
  \put(-260,123){(c)  ellipticity $e$}
\end{minipage}
\end{minipage}
\vspace*{-5pt}
\end{center}
\caption{\label{Fig:Skymaps_Scalar_Statistics}
  The aberrated temperature map is shown in panel (a),
  where a peculiar velocity of $\beta=0.01$ in the $z$-direction is used.
  The motion is thus towards the north pole (``above'') in this
  Mollweide projection.
  The panel (b) presents the eigenvalue $\lambda_1$
  defined in eq.(\ref{eq:lambda_1}).
  Finally, panel (c) displays the ellipticity $e$
  given in eq.(\ref{eq:ellipticity}).
  In order to enhance the visibility of the aberration effect,
  all three maps are smoothed in a final step by a
  Gaussian with a FWHM of $3^\circ$.
}
\end{figure}

In order to demonstrate that the twelve scalar measures presented in
section \ref{Sec:Scalars_Sphere} are not all suited on the same footing
to extract the modifications due to the aberration,
two examples of them are given in figure \ref{Fig:Skymaps_Scalar_Statistics},
where the velocity $\beta=0.01$ is used.
The panel \ref{Fig:Skymaps_Scalar_Statistics}(a) shows the aberrated
temperature map obtained from the CMB rest frame by applying the velocity
$\beta=0.01$ in the $z$-direction which corresponds to
the north pole (``above'') in these sky maps.
The aberration effect is almost invisible.
This changes drastically by computing the sky map which shows the value of
the first eigenvalue $\lambda_1$ of the Hessian matrix
instead of the aberrated temperature as revealed by
panel \ref{Fig:Skymaps_Scalar_Statistics}(b).
One recognizes a significant dipole contribution although the CMB rest frame
realization has none, since its quadrupole contribution is the first
non-vanishing multipole.
As an example from the group of distortion measures,
the figure \ref{Fig:Skymaps_Scalar_Statistics}(c) displays the
ellipticity $e$ defined in eq.(\ref{eq:ellipticity}), and
there is no asymmetry between the Northern and the Southern hemisphere,
which could reveal a non-vanishing peculiar velocity.
It should be noted that in all three panels of
figure \ref{Fig:Skymaps_Scalar_Statistics},
a Gaussian smoothing with a full width half-maximum (FWHM) of $3^\circ$ is
applied after the scalar measures are calculated from the expansion
coefficients $a^\text{(obs)}_{\ell m}$ of the unsmoothed temperature map
with $\ell_\text{max}=3000$.

\begin{figure*}
\begin{multicols}{2}
  \includegraphics[width=118pt]{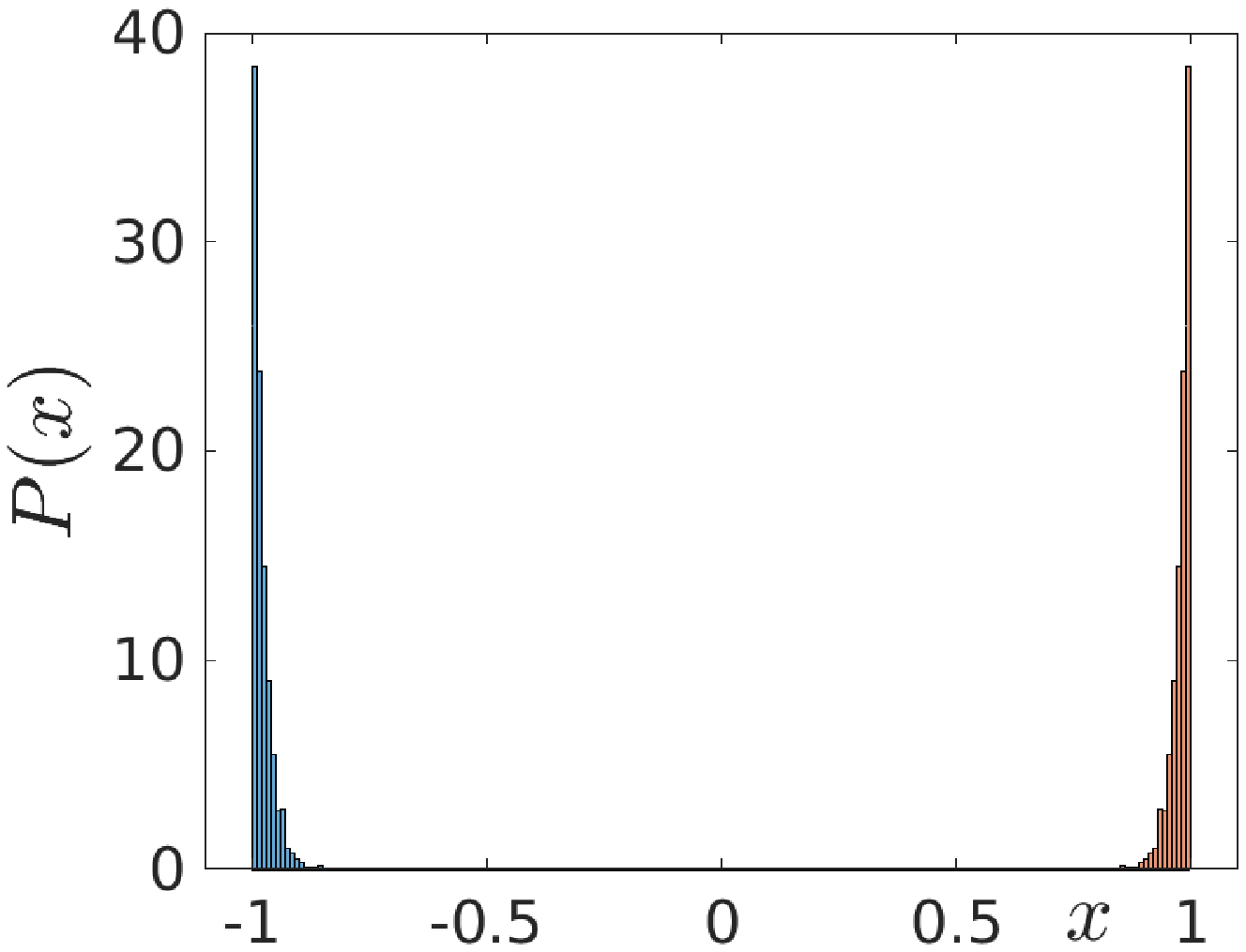}
  \includegraphics[width=118pt]{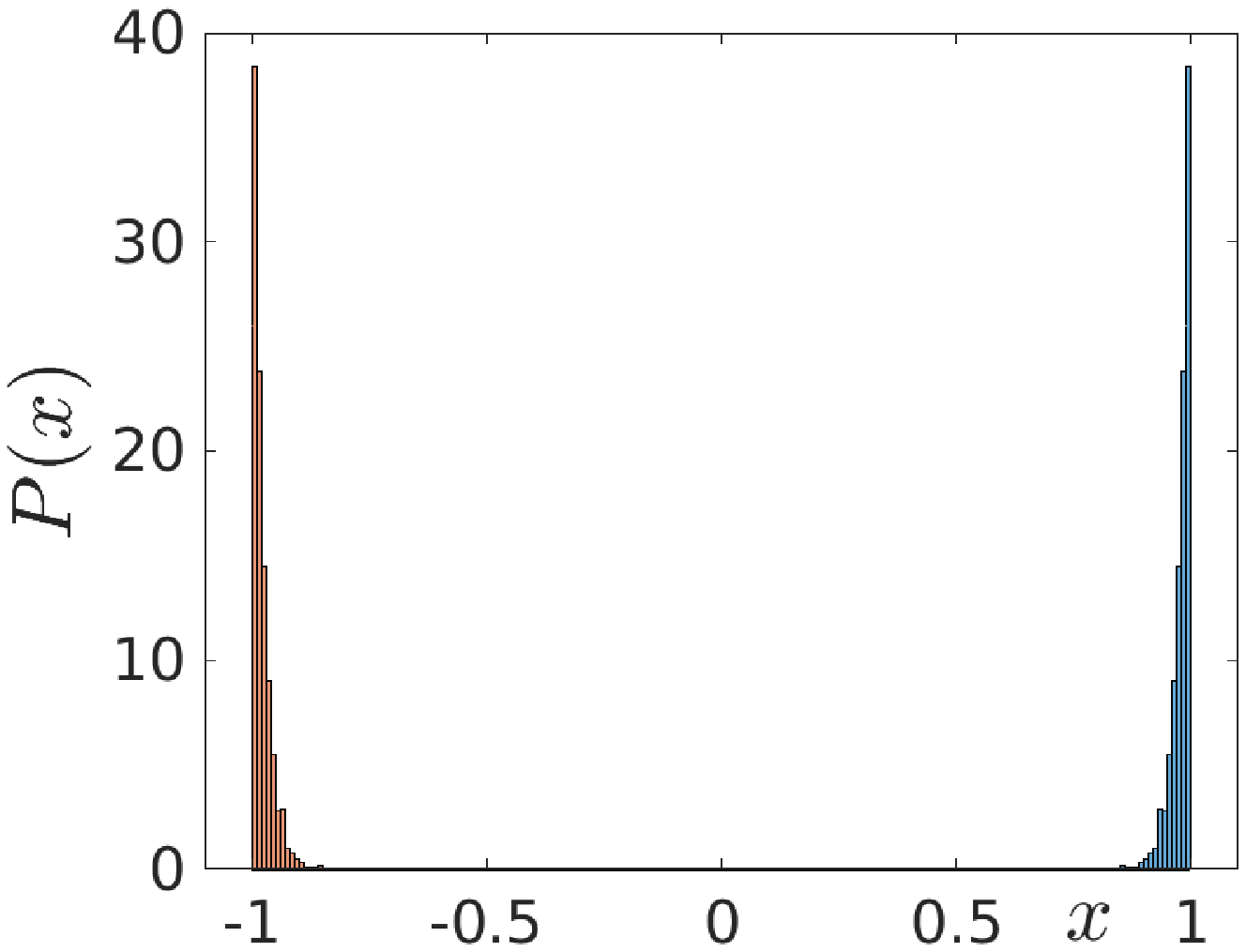}
  \put(-210,70){(a) $\lambda_1$}
  \put(-90,70){(b) $\lambda_2$}
  \par 
  \includegraphics[width=118pt]{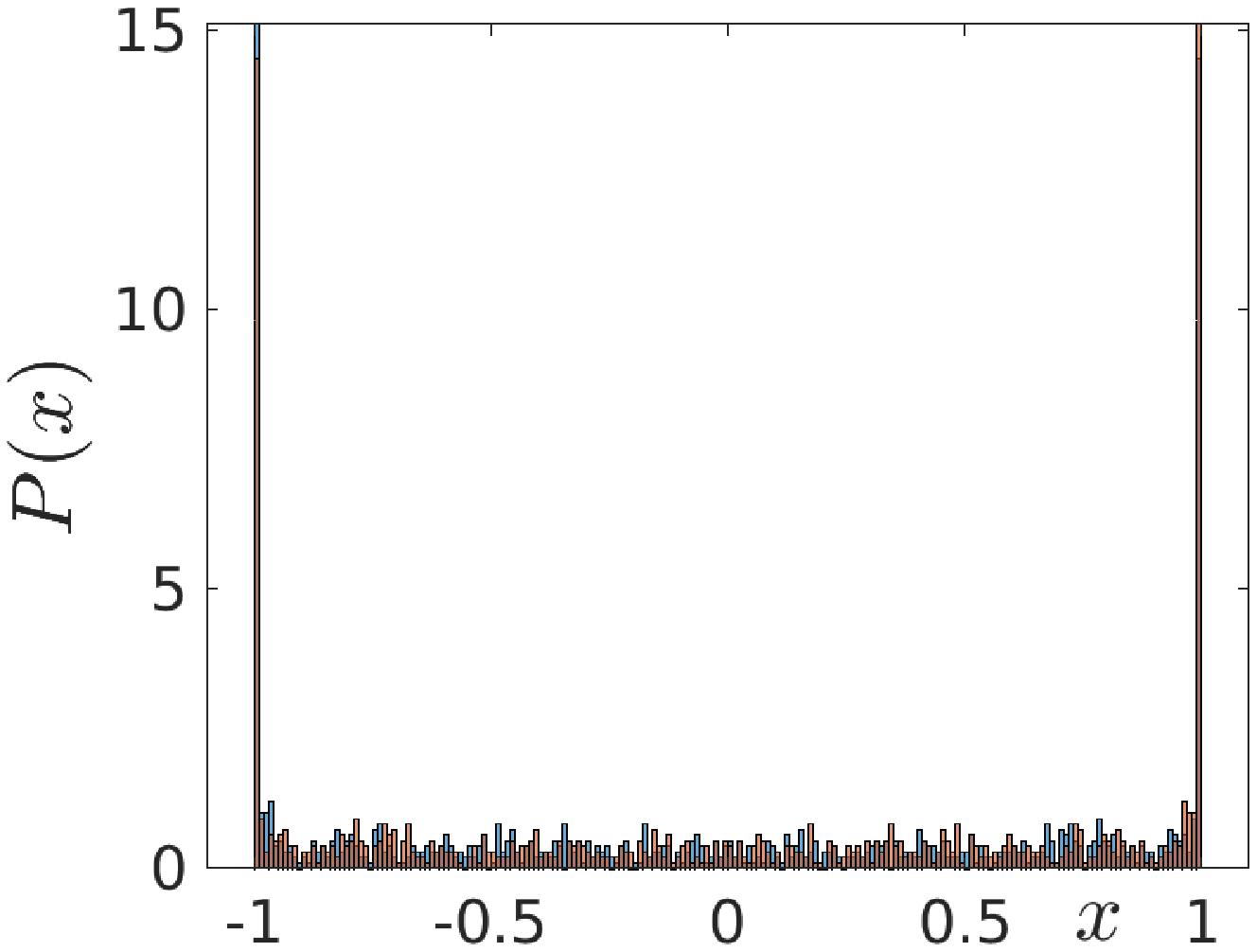}
  \includegraphics[width=118pt]{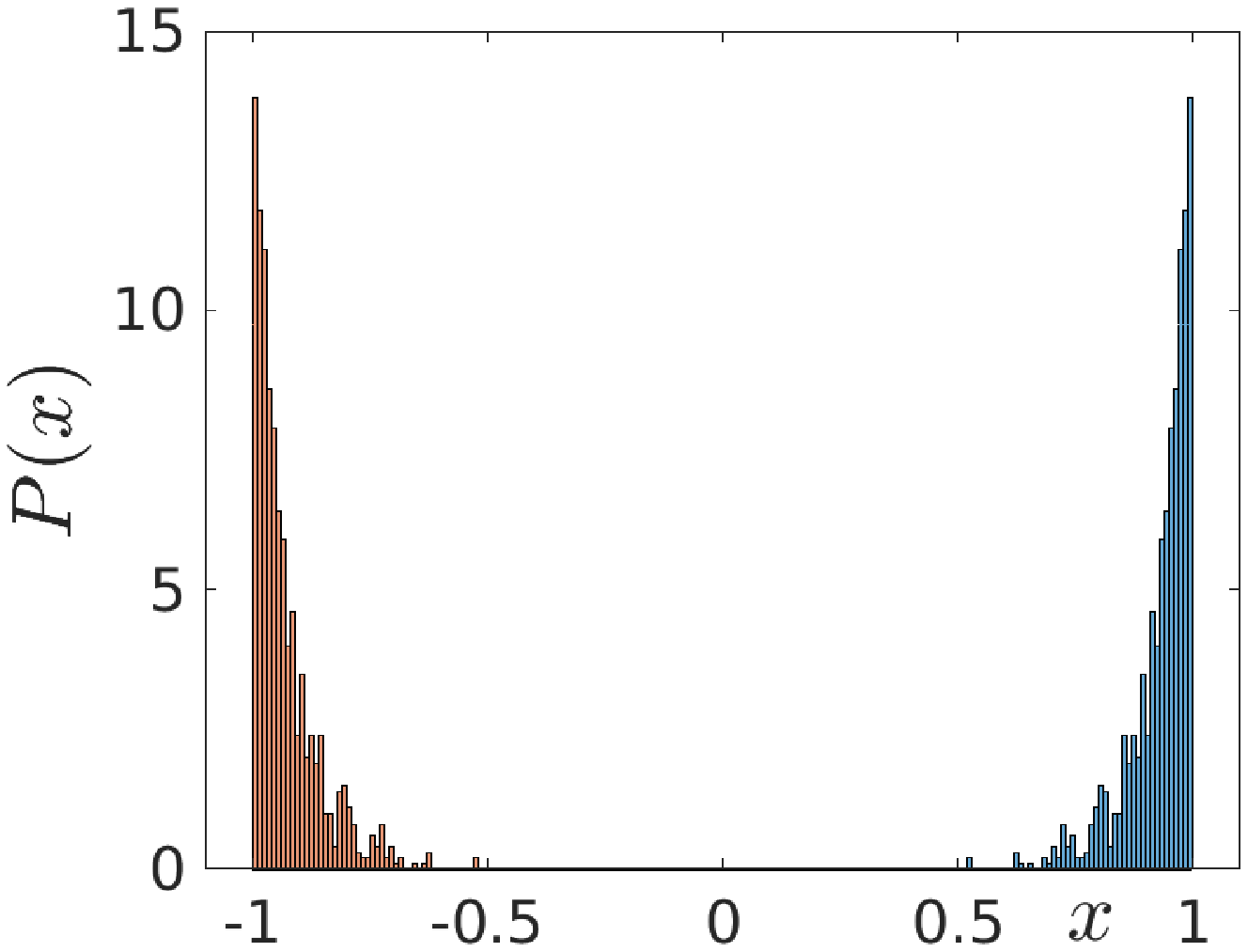}
  \put(-210,70){(c) Laplacian $\lambda_+$}
  \put(-90,70){(d) determinant $d$}
  \par
\end{multicols}
\begin{multicols}{2}
  \includegraphics[width=118pt]{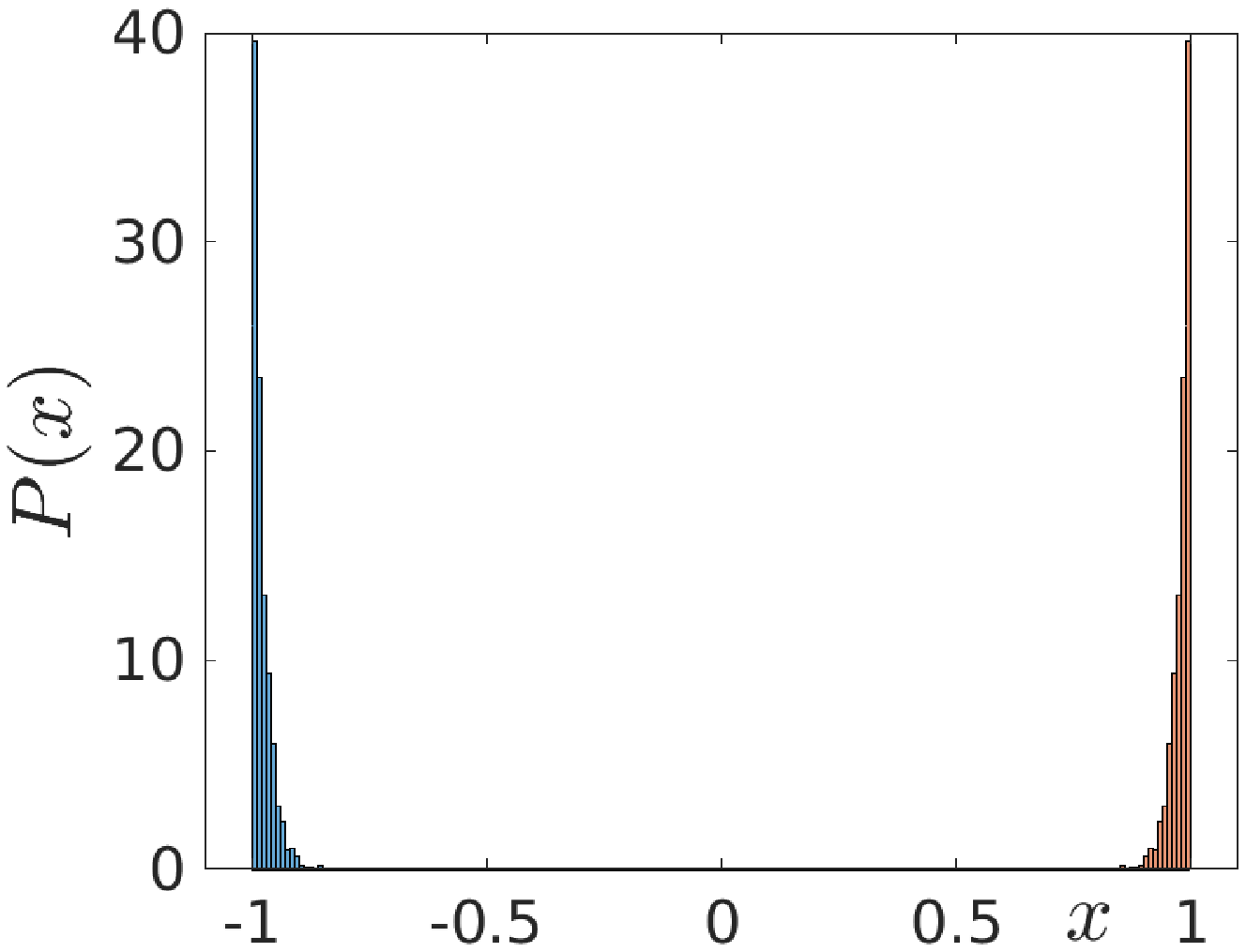}
  \includegraphics[width=118pt]{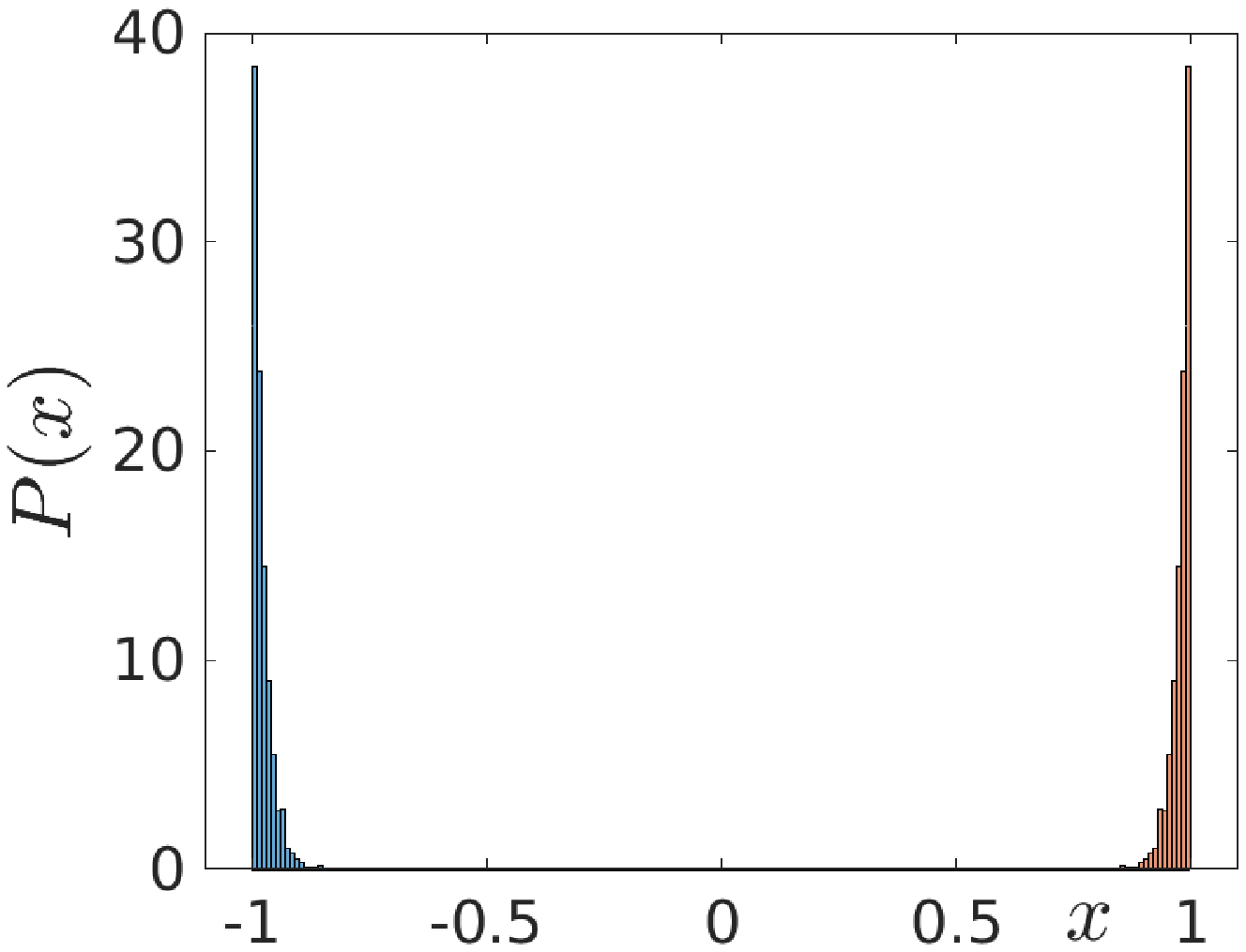}
  \put(-210,70){(e) shear $y$}
  \put(-90,70){(f) distortion $\lambda_-$}
  \par 
  \includegraphics[width=118pt]{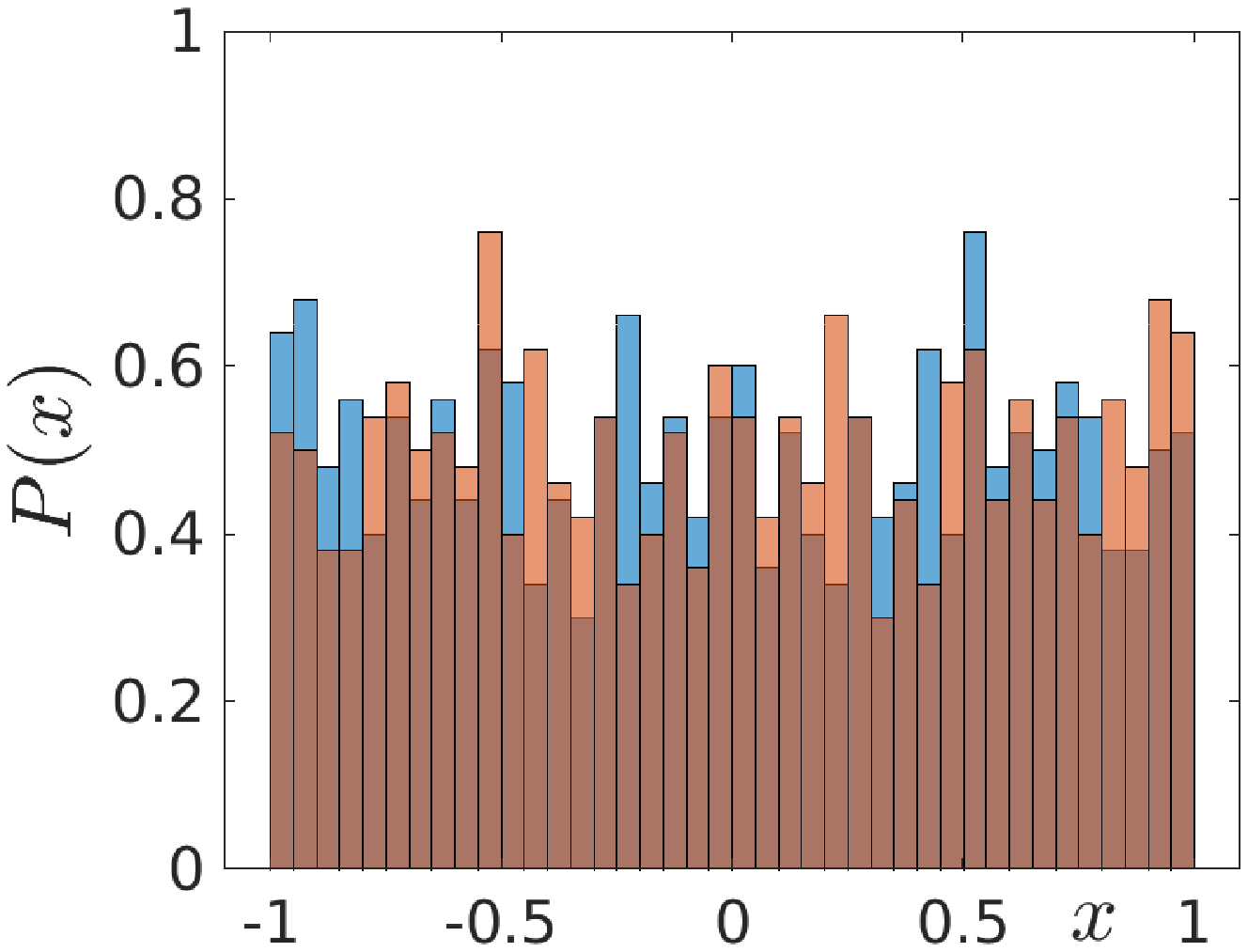}
  \includegraphics[width=118pt]{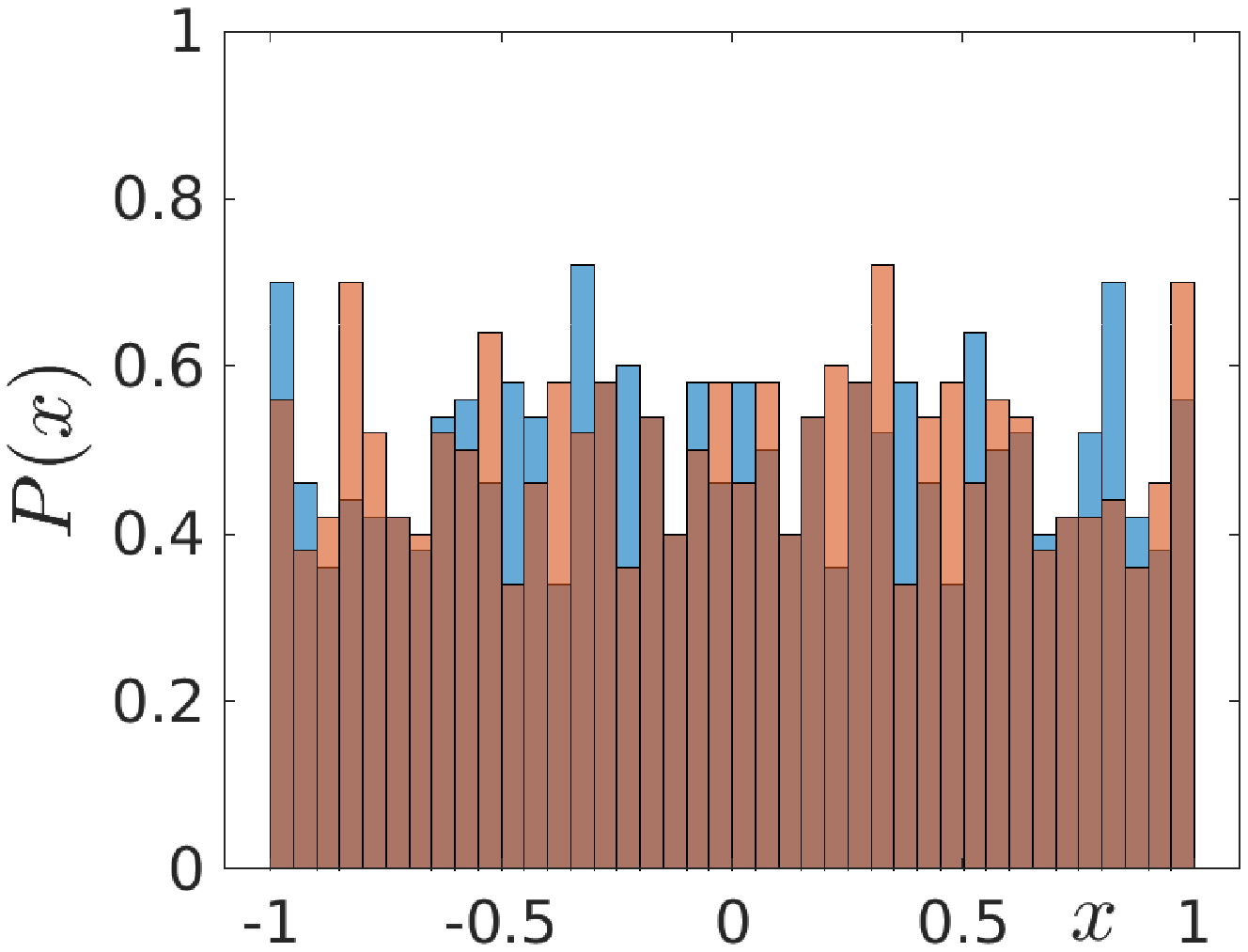}
  \put(-210,70){(g) ellipticity $e$}
  \put(-90,70){(h) shape index $\iota$}
  \par
\end{multicols}
\begin{multicols}{2}
  \includegraphics[width=118pt]{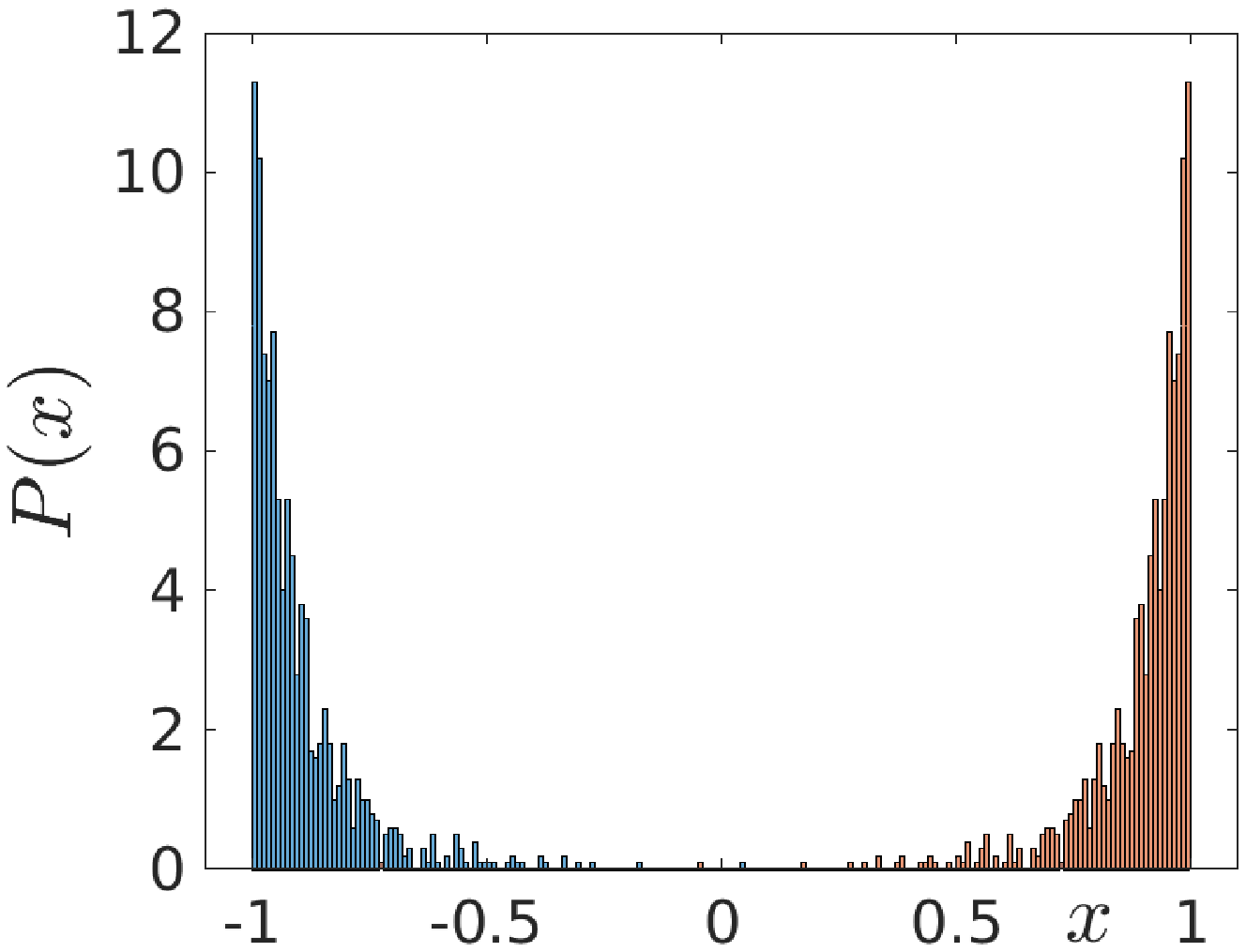}
  \includegraphics[width=118pt]{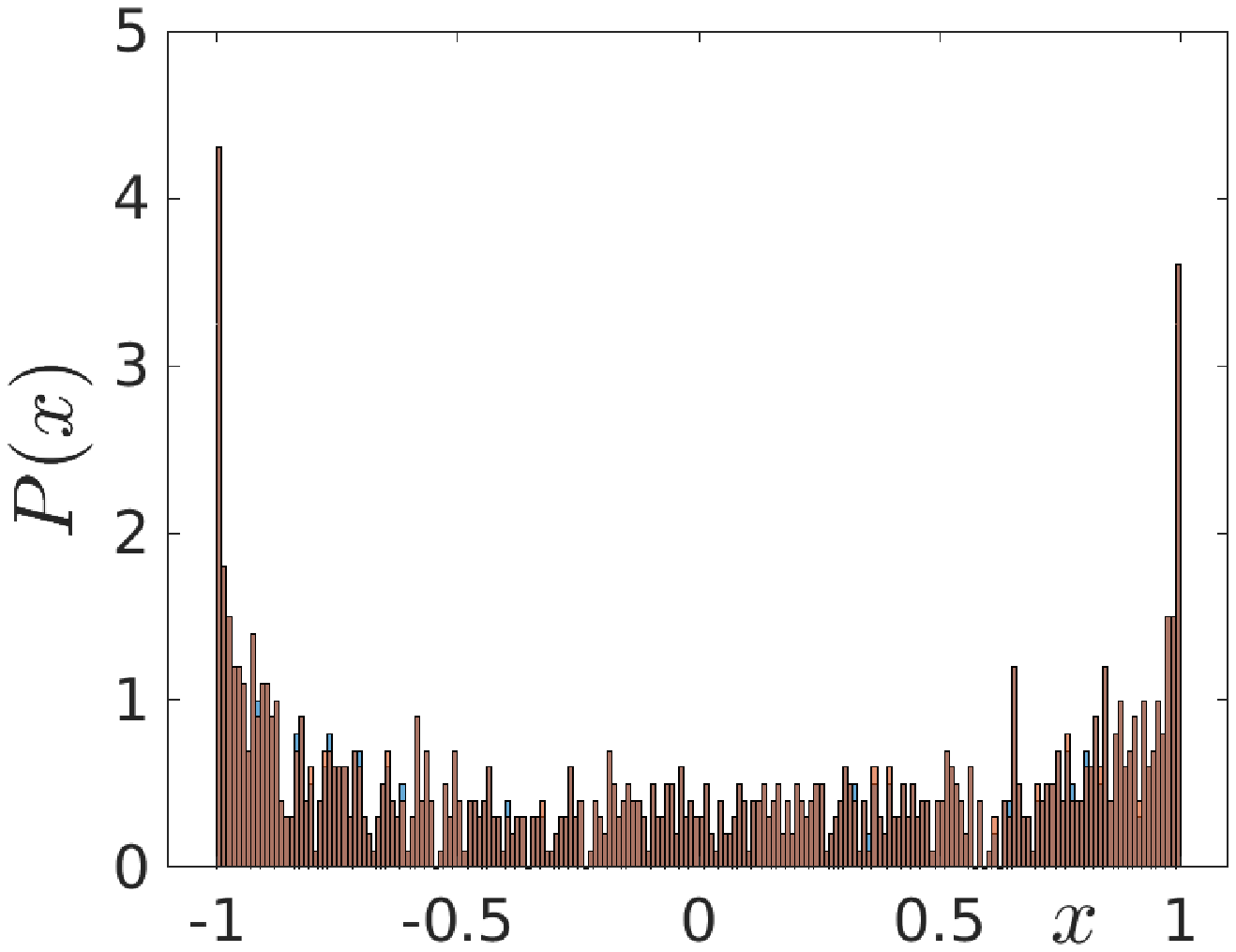}
  \put(-210,70){(i) $g$}
  \put(-90,70){(j) $D_g$}
  \par 
  \includegraphics[width=118pt]{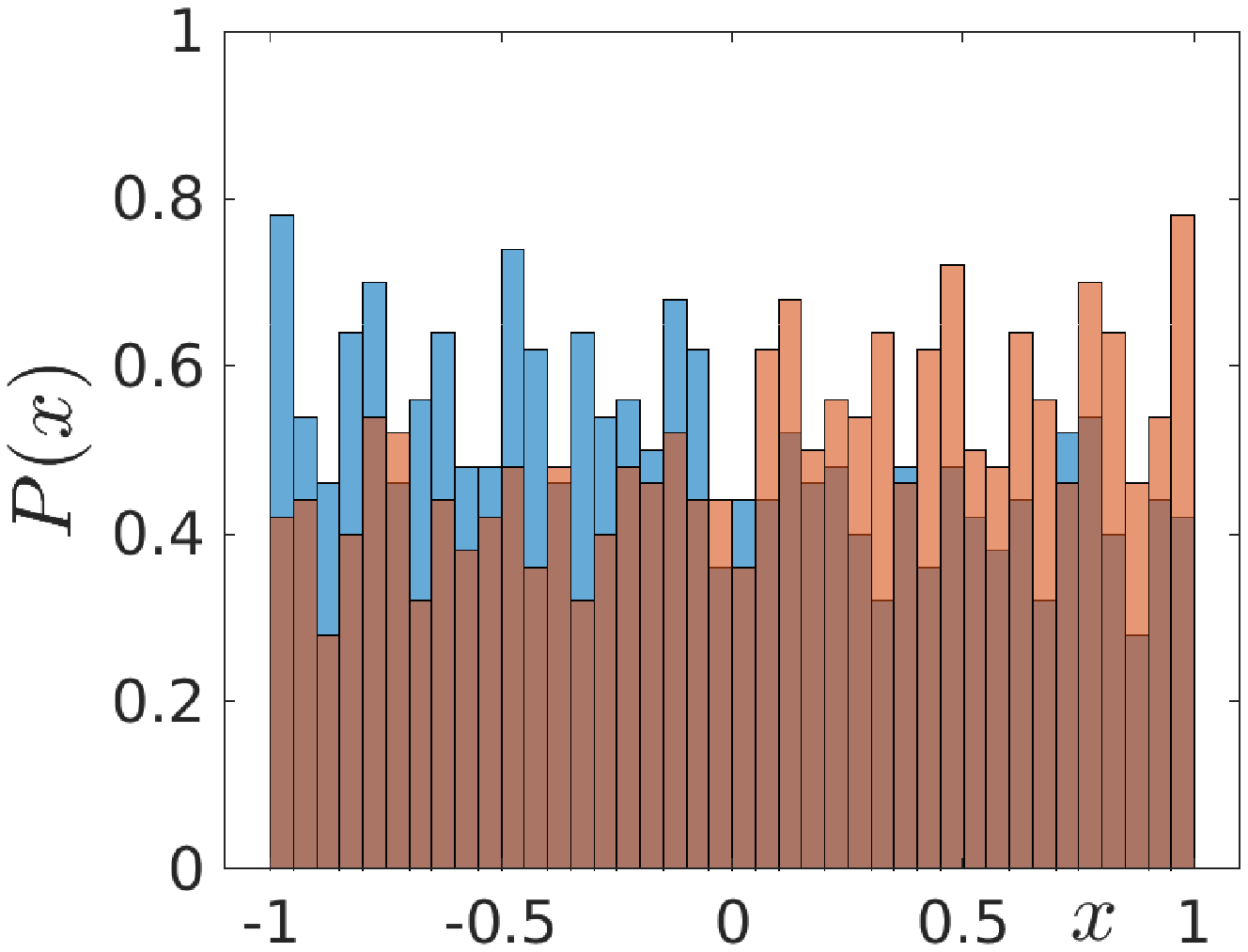}
  \includegraphics[width=118pt]{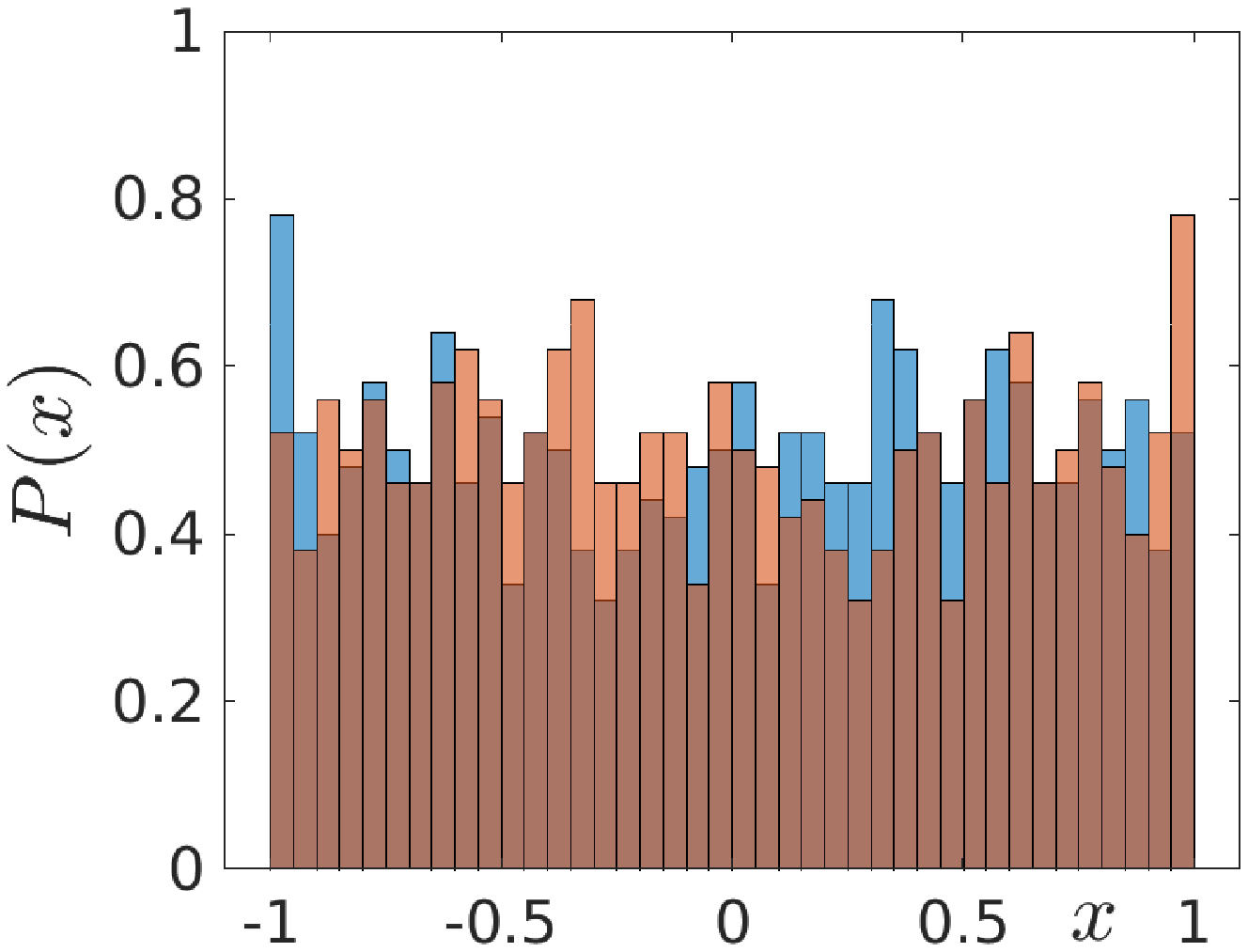}
  \put(-210,70){(k) $\kappa_\text{G}$}
  \put(-90,70){(l) $\kappa_\text{ex}$}
  \par
\end{multicols}
\caption{\label{Fig:histo_scalar_statistics}
  The probability distributions of the dipole orientations are shown obtained
  from 1000 CMB realizations for our solar system barycentre velocity
  $\beta_\odot=0.00123$.
  The bins have an equal width with respect  to $x=\cos\vartheta$ such that
  each bin corresponds to the same solid angle.
  The blue bins corresponds to the maximum of the dipole and the red ones
  to its minimum.
  The first row presents the distributions of the scalar measures derived
  from the Hessian matrix, while the second row displays the distortion
  scalars.
  The last row shows the square of the gradient modulus and its derivative
  as well as the two curvature measures.
  Four scalar measures possess an equidistribution.
  In these four cases, shown in panels (g), (h), (k), and (l),
  a larger bin width is chosen in order to suppress fluctuations.
}
\end{figure*}

Since figure \ref{Fig:Skymaps_Scalar_Statistics} has emphazised
that the scalar measures of section \ref{Sec:Scalars_Sphere} are
very different in their discriminating power to extract the aberration
effects, it is therefore natural to ask for the best-suited quantity.
To that aim a set of 1000 CMB realizations is generated using
the peculiar velocity $\beta_\odot= 0.00123$ and $\ell_\text{max}=3000$.
The boost is chosen in the $z$-direction and the twelve scalar measures
are computed from these 1000 aberrated CMB simulations.
In the next step, the dipole of each of the 1000 maps for each of the
twelve scalar measures is extracted
which allows the determination of the orientation of the dipole.
The agreement or the deviation of this orientation from the boost orientation
chosen in the simulations reveals the usability of the
considered scalar measure.
In spherical coordinates, the direction of the $z$-axis is given by
the polar angle $\vartheta=0$ which thus defines the correct forward direction.
The backward direction corresponds then to $\vartheta=\pi$, of course.
A suitable scalar measure should thus yield a value close to
$\vartheta=0$ for the forward direction and close to $\vartheta=\pi$
for the backward direction.
The distributions of the orientations obtained in this way are shown in
the histograms in figure \ref{Fig:histo_scalar_statistics}.
The bin width of histograms should be equal for the variable $x=\cos\vartheta$,
so that each bin refers to the same size of the corresponding solid angle.
Thus, the histograms discussed in the following cover the interval from
$\cos\vartheta=-1$ to $\cos\vartheta=1$.

The normalized histograms in figure \ref{Fig:histo_scalar_statistics} show the
probability distribution of $\cos\vartheta$ where the polar angle $\vartheta$
is determined from the orientation of the dipole $\ell=1$ of the
corresponding twelve scalar measures.
In order to compute this dipole orientation,
all expansion coefficients $a_{\ell m}$ are set to zero except those three
belonging to $\ell=1$ from which a sky map is generated
whose maximal and minimal values yield the dipole orientation.
Each panel in figure \ref{Fig:histo_scalar_statistics} displays two
histograms, one reveals the angle $\vartheta$
obtained from the maximum of the dipole and the other from the minimum.
As might be expected from figure \ref{Fig:Skymaps_Scalar_Statistics},
the eigenvalue $\lambda_1$ has a good discriminating power as revealed
by the sharp peaks in the histogram at $\cos\vartheta=-1$ and
$\cos\vartheta=1$.
The same applies to the other eigenvalue $\lambda_2$
although the orientation of the dipole is swapped.
The panel \ref{Fig:histo_scalar_statistics}(c) presents the histogram
of the Laplacian $\lambda_+$ given in eq.\,(\ref{eq:H_trace}).
Although the aberration direction is recognized in a lot of realizations,
there are also many false detections as revealed by the non-vanishing bins
even close to the equator.
This distracting feature is absent for the determinant $d$ of the
Hessian matrix, although the comparison of its histogram with
those of the eigenvalues $\lambda_1$ and $\lambda_2$ uncovers a
greater width and thus, a more unreliable detection of the boost direction.
The histograms in figures \ref{Fig:histo_scalar_statistics}(e) to 
\ref{Fig:histo_scalar_statistics}(h) corresponds to the group
of distortion measures.
The histograms reveal a good discriminating power for the shear $y$ and
for the distortion $\lambda_-$,
while the ellipticity $e$ and the shape index $\iota$ are unusable
as seen by the equidistributed histograms.
The sky map corresponding to the ellipticity $e$ was already shown
in figure \ref{Fig:Skymaps_Scalar_Statistics}(c),
where also no boost signature was noticeable.
The square of the gradient modulus $g$ is able to detect the boost direction
as seen in figure \ref{Fig:histo_scalar_statistics}(i),
but it is more unreliable than the eigenvalues $\lambda_1$ or $\lambda_2$
as one observes from the width of the histograms.
The derivative $D_g$ peaks also towards the peculiar velocity,
but it is indifferent with respect to its direction as seen in
figure \ref{Fig:histo_scalar_statistics}(j).
Finally, the last two panels reveal that the Gaussian curvature
$\kappa_\text{G}$ as well as the extrinsic curvature $\kappa_\text{ex}$
are unsuitable due to the equidistributed histograms.

Since there are four scalar measures with a good detection characteristic
as follows from the discussion of figure \ref{Fig:histo_scalar_statistics},
it is more quantitative to consider the median of their distributions.
The median values of the angular difference of direction of the dipole
from the boost direction is listed in table \ref{tab:scalar_measure_median}.
In the anti-boost direction, the listed value should be compared with
$\vartheta=180^\circ$.
One observes from table \ref{tab:scalar_measure_median}
that the best scalar measures are the two eigenvalues $\lambda_1$ and
$\lambda_2$, the shear $y$ and the distortion $\lambda_-$.
Although the square of the gradient modulus $g$ is more easily computed,
since only first order derivatives are needed, its median is roughly twice
as large as for the four best measures.
These median values are computed by using the solar system velocity and,
of course, would yield more accurate results for higher velocities.

\begin{figure}
\begin{center}
\hspace*{-10pt}\includegraphics[width=9.0cm]{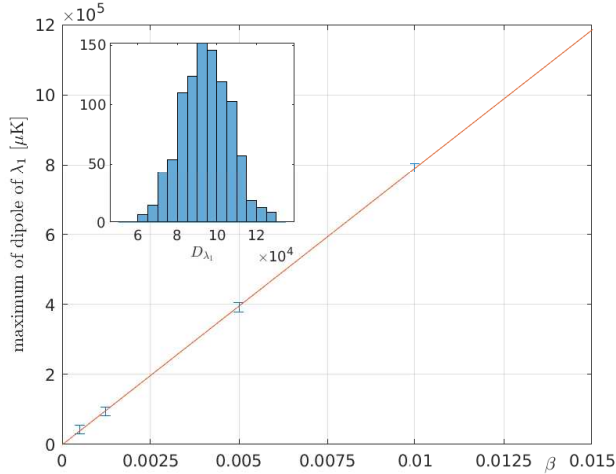}
\vspace*{-10pt}
\end{center}
\caption{\label{Fig:magnitude_lambda_1}
  The maximum $D_{\lambda_1}(\beta)$ of the dipole of $\lambda_1$ is plotted
  as a function of the peculiar velocity $\beta$.
  The error bars show the 1$\sigma$ deviation from the mean obtained
  from 1000 simulations for the values $\beta=0.0005, 0.00123, 0.005$
  and $0.01$.
  The linear fit given in eq.\,(\ref{eq:magnitude_dipol_lambda_1}) is shown
  as a straight line.
  The inset displays the distribution of $D_{\lambda_1}$ in units of $\mu\hbox{K}$ of the 1000
  realizations obtained for $\beta=\beta_\odot$.
}
\end{figure}

The eigenvalues $\lambda_1$ and $\lambda_2$ determine not only the direction
of the peculiar velocity but also its modulus.
This is demonstrated in figure \ref{Fig:magnitude_lambda_1}
where the maximal value $D_{\lambda_1}(\beta)$ of the dipole of $\lambda_1$
is plotted for several different values of the peculiar velocity $\beta$.
One observes a linear behaviour such that the amplitude of the dipole
can be used to estimate also the magnitude of the velocity.
The linear behaviour is well approximated by the fit
  \begin{equation}
  \label{eq:magnitude_dipol_lambda_1}
  D_{\lambda_1}(\beta) \, = \, c_1 \, \beta \, + \, c_0
  \end{equation}
  with
  \begin{equation}
  \label{eq:magnitude_dipol_lambda_1_coef}
  c_1 \, = \, (79.10 \pm 0.48) \times 10^6 \, \mu\hbox{K}
  \end{equation}
and
  \begin{equation}
  \label{eq:magnitude_dipol_lambda_0_coef}
  c_0 \, = \, - \, (783 \pm 2656) \, \mu\hbox{K}
  \hspace{10pt} ,
  \end{equation}
where the coefficients are determined from the simulations based on the
$\Lambda$CDM model specified above.
The inset in figure \ref{Fig:magnitude_lambda_1} displays the distribution
of the maximal values $D(\beta)$ for the 1000 simulations for the case
$\beta_\odot=0.00123$ which possess a standard deviation of
$\sigma=12699 \mu\hbox{K}$ around the mean value
$D_{\lambda_1}(\beta_\odot)=94391 \mu\hbox{K}$.
The standard deviation $\sigma$ is of the same order for the other considered
values of $\beta$.
Since the fit value for $c_0$ is compatible with zero,
the constant $c_0$ can be neglected. 
Thus, if the observations would determine $D_{\lambda_1}$ with
sufficient accuracy,
the modulus of velocity $\beta$ could be determined with an accuracy
of the order $\sigma/c_1 \simeq 1.6 \times 10^{-4}$.

\begin{table}
  \caption{
    The median values of the polar angle $\vartheta$ of the distributions
    of the suitable scalar measures are listed,
    such that in 50\% of the cases,
    a more accurate estimate of the velocity would be obtained.
    The median values are derived from 1000 aberrated CMB realizations
    with $\beta_\odot = 0.00123$.
    Since some scalar measures have their minimum of the dipole in
    the forward direction instead of the backward direction,
    the median direction of the minimum is additionally listed
    for convenience, which is related by
    $\vartheta_\text{min} = 180^\circ-\vartheta_\text{max}$.
  }
 \label{tab:scalar_measure_median}
 \begin{tabular*}{\columnwidth}{@{}l@{\hspace*{30pt}}l@{\hspace*{30pt}}l@{}}
  \hline
  Scalar measure & Median of        & Median of     \\
                 & dipole maximum   & dipole minimum \\
  \hline
  $\lambda_1$ & 170.5324$^\circ$ & 9.4676$^\circ$ \\
  $\lambda_2$ & 9.4676$^\circ$ & 170.5324$^\circ$ \\
  $\lambda_+$ & 91.4549$^\circ$ & 88.5451$^\circ$ \\
  $d$         & 17.416$^\circ$ & 162.584$^\circ$ \\
  \hline
  $y$         & 170.498$^\circ$ & 9.502$^\circ$ \\
  $\lambda_-$ & 170.5324$^\circ$ & 9.4676$^\circ$ \\
  \hline
  $g$         & 159.6431$^\circ$ & 20.3569$^\circ$ \\  
  \hline
 \end{tabular*}
\end{table}

\begin{figure}
\begin{center}
\hspace*{-10pt}\includegraphics[width=9.0cm]{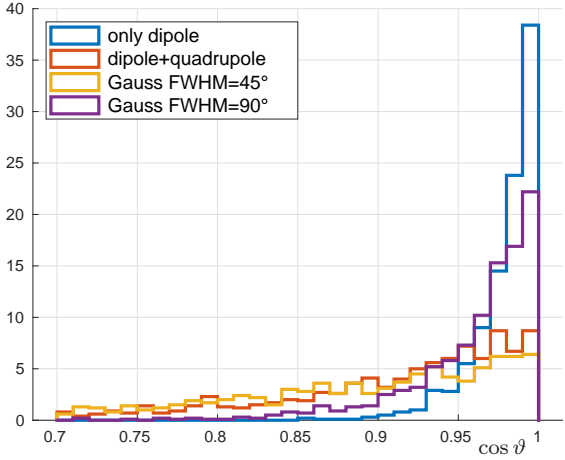}
\put(-56,8){$\cos\vartheta$}
\vspace*{-10pt}
\end{center}
\caption{\label{Fig:compare_dipole_gauss}
  The distribution of the deviation $\cos\vartheta$ is determined from
  the sky map of the first eigenvalue $\lambda_1$.
  The deviation angle $\vartheta$ is determined in three ways:
  (i) the orientation of the dipole by its minimum on the sphere,
  (ii) the minimum of the sum of the dipole and the quadrupole, and
  (iii) the minimum of the sky map obtained from a Gaussian smoothing
  with a FWHM=$45^\circ$ and with a FWHM=$90^\circ$ but retaining all multipoles.
}
\end{figure}

Up to now, the peculiar velocity is solely determined from the dipole
orientation extracted from the corresponding scalar measure.
Thus, one might wonder whether the inclusion of higher multipoles
could increase the accuracy.
To that aim, the extrema in the sky maps computed from the sum of
the dipole and the quadrupole contribution are also investigated.
Because one might also worry about the sharp cut-off at $\ell=2$ or $\ell=3$,
respectively, a strong smoothing with Gaussian of FWHM of $45^\circ$
and of $90^\circ$ is additionally taken into account.
The results are presented in figure \ref{Fig:compare_dipole_gauss} for
the example of the first eigenvalue $\lambda_1$.
The position of the minimum of the values of $\lambda_1$ on the sphere
are determined,
which should point approximately in the boost direction $\vartheta=0$.
It is clearly seen that the sharpest distribution towards
$\cos\vartheta=1$ is obtained by using only the dipole.
Taking higher multipoles into account leads to a smearing of the
distribution and thus to a more unreliable estimate for the boost direction.

In the case of partial sky observations, one has to deal with the difficulty
to determine the dipole in the presence of a masked sky.
The dipole then has to be fitted to the map of the eigenvalue $\lambda_1$
in the observed part of the sky.
The full sky analysis of this work leads to the suggestion
that the best estimate for our peculiar velocity will probably be achieved
by fitting the dipole to the unsmoothed map.
Without smoothing one does not need to care about smoothing a masked sky
which is usually carried out in the $a_{\ell m}$ space by multiplying
the expansion coefficients $a_{\ell m}$ with
$F_\ell = \exp(-\frac 12\alpha^2\theta_G^2\,\ell(\ell+1))$ and
$\alpha = \pi/(180\sqrt{8\ln 2})$
if the smoothing $\theta_G$ is given in degrees.
This requires the determination of the coefficients $a_{\ell m}$ from a
partial sky observation with a probably complicated mask structure.
We thus suggest a three parameter fit,
which is the dipole amplitude and the two angles for its orientation,
to the unsmoothed map using a least-squares fitting, for example.
This can be achieved by minimizing the expression
\begin{equation}
\label{eq:minimizing}
S(\beta,\vartheta,\varphi) \, = \,
\sum_i \frac{(p^i_\text{dip}(\beta,\vartheta,\varphi)-p^i)^2}{\sigma_i^2}
\hspace{10pt} ,
\end{equation}
where $p^i_\text{dip}(\beta,\vartheta,\varphi)$ denotes the dipole value
at pixel $i$ with the dipole orientation given by $\vartheta$ and $\varphi$.
Furthermore, $p^i$ is the value of the $\lambda_1$ map at pixel $i$
and $\sigma_i$ the estimated error for this pixel.
The sum runs only over the observed pixels.
This procedure should also be favourable to reduce the uncorrelated
pixel-to-pixel noise.
It is clear that observations of too small a fraction of the sky
will not allow a reliable determination of the peculiar velocity.

\section{Conclusion}

The motion of our solar system relative to the CMB rest frame leads to
the large CMB dipole and to subtle distortions in the CMB sky map
due to the aberration effect.
Our peculiar velocity is usually derived from the dipole component of
the CMB sky.
It is then assumed that the CMB dipole is a pure Doppler dipole
and a further intrinsic dipole and secondary effects can be safely ignored.
It is thus desirable to have an independent method for the determination
of the peculiar velocity.
The aberration effect leads to mode couplings in the angular power spectrum
which can betray the motion as discussed in the Introduction.
However, if one tries to circumvent the determination
of the spherical harmonic expansion coefficients $a_{\ell m}$,
one can consider covariant derivatives on the sphere
which measure the extent of the distortions
being a compression of structures in the forward direction and
a stretching in the backward direction.
This method thus requires CMB maps with sufficiently high resolution.
But since no full sky expansion is required as long as the process of
derivation can be carried out with the observational data,
high-resolution partial data would be suitable as provided by
the Simons Observatory \citep{Ade_et_al_2019}
and by CMB-S4 \citep{Abazajian_et_al_2019}.
The direct use of the partial derivatives on the sphere with respect to
the coordinates $\vartheta$ and $\varphi$ has the disadvantage
that they depend on the orientation of the spherical coordinate system.
This is circumvented by the use of scalar measures,
where in this work the twelve scalar quantities suggested by
\cite{Monteserin_Barreiro_Sanz_Martinez-Gonzalez_2005}
are investigated.
The main result is that the eigenvalues $\lambda_1$ and $\lambda_2$ of the
Hessian matrix defined in eqs.(\ref{eq:lambda_1}) and (\ref{eq:lambda_2}),
the shear $y$ and the distortion $\lambda_-$ given in
(\ref{eq:shear}) and (\ref{eq:distortion}), respectively,
are the best choices for detecting the boost direction.
The remaining eight scalar quantities are not well-suited for this task.
And it is demonstrated that the peculiar velocity is most accurately
determined by extracting only the dipole contribution from the four
favourable scalar measures.


\section*{Acknowledgements}

The software packages HEALPix (\url{http://healpix.jpl.nasa.gov},
\cite{Gorski_Hivon_Banday_Wandelt_Hansen_Reinecke_Bartelmann_2005})
and CAMB written by A.~Lewis and A.~Challinor (\url{http://camb.info/})
as well as the Planck data from the LAMBDA website
(\url{http://lambda.gsfc.nasa.gov}) were used in this work.
Furthermore, we would like to thank the unknown referee for his useful
comments.


\section*{DATA AVAILABILITY}

The data underlying this article will be shared on reasonable request to
the corresponding author.


\bibliography{../../bib_astro.bib}
\bibliographystyle{mnras}


\label{lastpage}

\end{document}